%% file: thesis.tex
\documentclass[titlepage,12pt,a4paper]{report}

\usepackage{thesis}

\begin{document}

%\maketitle

\begin{titlepage}
  \addtolength{\oddsidemargin}{-1.35cm}
  \addtolength{\linewidth}{3.2cm}

  \vspace*{-1cm}
  \rule{\linewidth}{2pt}\\[10pt]
  \begin{minipage}{\linewidth-0.4cm}
    \begin{flushright}
      \usefont{T1}{phv}{m}{n}\fontsize{30pt}{30pt}\selectfont
      The polygon model for 2+1D gravity:\\[10pt]
      the constraint algebra and\\[10pt]
      problems of quantization
    \end{flushright}
  \end{minipage}\\[10pt]
  \rule{\linewidth}{2pt}

  \vspace*{\stretch{1}}

  \begin{minipage}{\linewidth}
    \centering\large

    Jaap Eldering\\[15pt]

    7th June 2006\\[15pt]

    Masters thesis\\
    supervised by Prof. Renate Loll\\
    Institute for Theoretical Physics\\
    Utrecht University\\
  \end{minipage}

  \vspace*{\stretch{2}}

  \raisebox{-1.5cm}[0pt][0pt]{\includegraphics{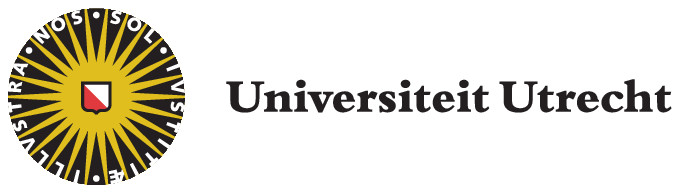}}
  
\end{titlepage}

\include{abstract}

\tableofcontents

\include{introduction}
\include{notation}
\include{representation}
\include{polygon}
\include{quantization}
\include{constraintalgebra}
\include{conclusions}

\appendix
\include{dualgraph}

\include{acknowledgements}

% Bibliography stuff:
\addcontentsline{toc}{chapter}{Bibliography}
\bibliographystyle{thesis}
\bibliography{bibfile}

\end{document}

%% file: abstract.tex
\begin{abstract}
  In this thesis we consider 't Hooft's polygon model for 2+1D
  gravity. We first recall the ADM formalism to write general
  relativity in Hamiltonian form. With this background in mind, we
  give a detailed review of the polygon model and its explicit
  evolution description in the classical context.

  Then we review some remarks in the literature about quantization of
  this model and the discreteness of space-time. We discuss the
  problems associated with this in the context of canonical
  quantization for some explicit quantization schemes: the triangle
  inequalities conflict with the Stone--Von Neumann uniqueness theorem
  when we use the canonical variables. Also, the implementation of
  transitions, the constraints and their algebra poses significant
  problems. We conclude that no rigorous conclusions about the
  spectrum of space-time can be drawn without an explicit quantization
  scheme.

  Furthermore, we consider the Poisson structure of the constraints,
  which are important when we try to quantize the model. We improve
  known results and show that the full Poisson structure can be
  calculated explicitly and that it closes on shell. An attempt is
  made to interpret the gauge orbits generated by the constraints.
\end{abstract}

%%% Local Variables: 
%%% mode: latex
%%% TeX-master: "thesis"
%%% End: 

%% file: introduction.tex
\chapter{Introduction}
\label{chap:introduction}

In this thesis we will study a particular model for (Einstein)
gravity: the 't Hooft polygon model for gravity in $2\tp1$\ndash
dimensions. We will specifically be concerned with the question of
whether this model can be quantized and we calculate the complete
Poisson structure of the constraints.

The $2\tp1$ here stands for 2 spatial dimensions and 1 time dimension.
This is one dimension lower than the $3\tp1$\ndash dimensions we
(seem to) live in. That turns out to make the whole problem a lot
simpler than trying to tackle the full problem of four-dimensional
quantum gravity, which people are trying to solve for decades already.
It is hoped then, that by studying and hopefully solving
three-dimensional (and in general lower dimensional) quantum gravity,
we can learn something about how to solve four-dimensional quantum
gravity.

\section{Gravity: general relativity}

When we say `gravity' we mean the classical theory of gravity as
formulated by Albert Einstein, which is known as `general relativity'.
It has as a starting point that (locally) the effect of a
gravitational field and acceleration are the same.

This has as consequences that space and time together (often referred
to as space-time) can be curved: the concept of a straight line is not
entirely lost, but becomes somewhat more subtle. For example in a
curved space it can be possible to always travel in a straight line
and come back to where you started, or that two straight lines
intersect each other at more than one point.

The mathematical description of this theory is much more complicated
than that of the classical, Newtonian theory of gravity and finding
solutions is much more difficult, already in the classical
(non-quantum) theory.

On the other hand, general relativity describes some interesting
phenomena. Some of these are already well confirmed by measurements,
like the bending of light rays by large masses (e.g. the sun) and the
slowing of time in gravitational wells. Other phenomena are predicted
but not yet detected, like black holes and gravitational waves.

\section{Quantization}

At the start of the 20th century, physical phenomena were discovered
which could not be described by the classical physical theories.
Examples are the photo-electric effect and the observed structure of
atoms.

A theory which could correctly describe all of these phenomena was
constructed over the years and is generically called `quantum
mechanics'. In this theory, observables like particle position or
energy do not have definite values anymore; these observables have a
probability-like amplitude. The `quantum' refers to the fact that
certain observables can only take (a superposition of) discrete,
quantized values in this theory.

At this moment, a theory for three out of four known fundamental
forces in nature has been formulated: the electro-magnetic force and
the weak and strong nuclear forces. Only the force of gravity remains
``unquantized'' as of this day. Yet one cannot consistently combine
classical and quantum theories, so the current theories of general
relativity and quantum mechanics are not complete. Hence physicists
are looking for a quantized version of the theory of gravity. For more
details on the subject of quantization, we refer to
\fref[plain]{chap:quantization}.

\section{The polygon model}

In $3$ dimensions, for solutions of the Einstein equations without
cosmological constant, curvature of space-time does only occur locally
at points where there is mass. Everywhere where space is empty, it is
(locally) flat. This doesn't mean however that things become trivial,
because we can still have global non-trivial curvature effects.

The polygon model for $2\tp1$\ndash dimensional gravity makes clever
use of this fact. We can model the spatial part of space-time as a set
of polygons glued together. Inside each polygon space is simply flat,
but these polygons can be glued together to form a two-dimensional
spatial surface with non-vanishing two-dimensional curvature and
non-trivial topology. We could for example create a cube from six
four-sided polygons. The polygon model is explained in more detail in
\fref[plain]{chap:polygon}.

There are other models for $2\tp1$\ndash dimensional gravity. Two
formulations that are closely related to each other and the polygon
formulation are the second order ADM formulation with York time by
Moncrief et al. \cite{Moncrief:1989dx,Carlip:1998uc} and the first
order Chern-Simons formulation by Witten et al.
\cite{Witten:1988hc,Carlip:1998uc}. On a classical level, these
formulations are equivalent. The quantizations of these formulations,
as far as they exist, yield different results though. Witten's
formulation gives a `frozen time' picture and Moncrief's formulation
can only be explicitly quantized in the genus 1 case.

%%% Local Variables: 
%%% mode: latex
%%% TeX-master: "thesis"
%%% End: 

%% file: notation.tex
\chapter{Notation and conventions}
\label{chap:notation}

In the polygon model, the fundamental variables are the lengths and
boost parameters of the edges between polygons. These polygons cover a
spatial slice of space-time. Boost parameters are the generators of
Lorentz transformations between different polygons. Explicitly a boost
$\eta$ is related to the speed $v$ of a Lorentz transformation by
\begin{equation}
  \label{eq:notation:hyperbolic-boost}
  \begin{split}
    \tanh(\eta) &= v,\\
    \cosh(\eta) &= \gamma(v),\\
    \sinh(\eta) &= \gamma(v) \, v,
  \end{split}
\end{equation}
where as usual we have
$\gamma(v) = \dfrac{1}{\sqrt{1-\left(\mfrac{v}{c}\right)^2}}$ and we
normalize $c = 1$.

The boost parameters of edges can be viewed as Lorentz transformations
in the specific direction perpendicular to that edge. We can more
generally consider boosts in all directions together with spatial
rotations. These form the (Lie) group of Lorentz transformations: we
can compose any two elements to get another Lorentz transformation. We
write $O(2,1)$ for the group of Lorentz transformations in
three-dimensional space-time. This group can be generated\footnote{%
  Actually this will only generate the identity component
  $SO^+(2,1) \subseteq O(2,1)$, but this will be sufficient for our
  purposes; see appendix~\fref[plain*]{chap:dualgraph} for more
  details.} %
by the three basic transformations of a rotation and boosts in the $x$
and $y$\ndash direction:
\begin{equation}
\label{eq:notation:lorentzmatrices}
\begin{split}
  L_r(\alpha) &= \left(\begin{array}{ccc}
      1 & 0 & 0\\
      0 & \cos\alpha &-\sin\alpha\\
      0 & \sin\alpha & \cos\alpha
    \end{array}\right),\\
  L_x(\eta) &= \left(\begin{array}{ccc}
     \cosh\eta & \sinh\eta & 0\\
     \sinh\eta & \cosh\eta & 0\\
      0 & 0 & 1
    \end{array}\right),\\
  L_y(\eta) &= \left(\begin{array}{ccc}
      \cosh\eta & 0 & \sinh\eta\\
      0 & 1 & 0\\
      \sinh\eta & 0 & \cosh\eta
    \end{array}\right).
\end{split}
\end{equation}
As generators of Lorentz transformations, boosts have the nice property
of additivity:
\begin{equation}
  L_x(\eta_1)\,L_x(\eta_2) = L_x(\eta_1+\eta_2).
\end{equation}

We will denote boosts by Greek letters from the middle of the
alphabet ($\eta, \mu, \nu, \dots$), angles by Greek letters from the
start ($\alpha, \beta, \gamma, \dots$) and lengths by $l, m, n, \dots$

In the polygon model, we will frequently encounter hyperbolic
functions of the boosts and trigonometric functions of angles. To keep
formulas short and clear, the following notation will be used:
\begin{equation}
  \label{eq:notation:boost-shorthand}
  \begin{split}
  v_i      &\equiv v_{\eta_i}      \equiv \tanh(\eta_i),\\
  \gamma_i &\equiv \gamma_{\eta_i} \equiv \cosh(2\,\eta_i),\\
  \sigma_i &\equiv \sigma_{\eta_i} \equiv \sinh(2\,\eta_i),\\
  c_i      &\equiv c_{\alpha_i}    \equiv \cos(\alpha_i),\\
  s_i      &\equiv s_{\alpha_i}    \equiv \sin(\alpha_i).
  \end{split}
\end{equation}
Note the extra factor 2 in the definition of $\gamma$ and $\sigma$.
Furthermore, when the specification of a variable is obvious from the
subscript index, then this shorter notation will be used.

\begin{figure}[b]
  \centering
  \begin{minipage}[t]{.33\textwidth}
    \centering
    \includegraphics{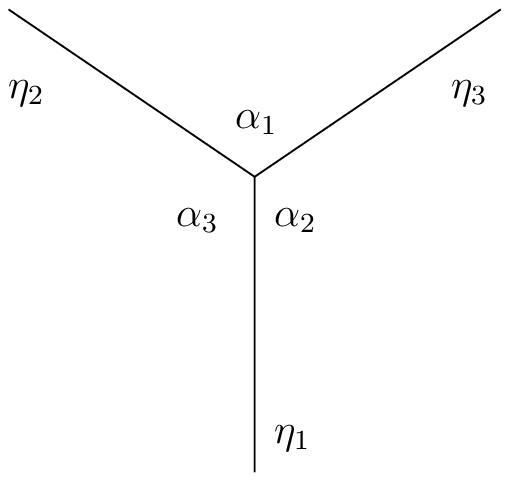}
    \caption{Labels of edges and angles at a vertex.}
    \label{fig:vertex-params}
  \end{minipage}
  \hspace{1.5cm}
  \begin{minipage}[t]{.50\textwidth}
    \centering
    \includegraphics{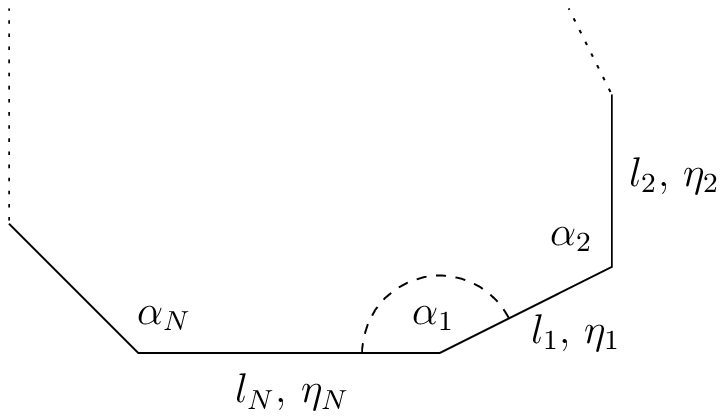}
    \caption{Labels of edges and angles along a polygon.}
    \label{fig:polygon-params}
  \end{minipage}
\end{figure}
Boosts, angles and lengths will sometimes be indexed along the border
of a polygon and sometimes along a vertex. To distinguish between
those, we will label the former with indices $i,j,k$ and the latter
with indices $a,b,c$. Boosts and angles with the same index are
related in different ways for those cases. At a vertex, a boost and
its opposite angle have the same index, while along a polygon an angle
and the boost of the counterclockwise following edge have the same
index. See \fref[plains]{fig:vertex-params} and
\fref[plain*]{fig:polygon-params}. Note that when edges and/or
vertices have to be numbered, this will always be in a
counterclockwise orientation.

Furthermore we see that indices along a polygon are cyclic. In
computations we will therefore always implicitly interpret the index
numbers modulo $N$ ($N$ being the number of vertices and edges of the
polygon). For example, we have that $\eta_0 = \eta_N$. An exception to
this rule will be made for quantities that are sums over ranges of
indices, as a sum over the empty range is truly different from a sum
over the full cycle. For example, the sum of angles up to index $j$ as
defined in~\fref{eq:polygon:theta}: here we interpret
\begin{equation}
  \label{eq:notation:cyclicindex}
  \theta_0 = \sum_{i=1}^0 \pi - \alpha_i = 0
\end{equation}
instead of $\theta_0 = \theta_N$.

The model has a phase space consisting of a set of edge lengths and
boosts $(l_i,\eta_j)$. The Poisson brackets are defined on this phase
space as
\def\poissonleta{\tfrac{1}{2}\,\delta_{ij}}
\begin{equation}
  \label{eq:notation:poisson}
  \poisson{l_i}{\eta_j} = \poissonleta,
\end{equation}
where the indices $i,j$ run over all edges.

The phase space of this model is larger than the physical phase space.
This is reflected by the presence of algebraic constraints on the
phase space variables. These define a subspace of the complete phase
space by imposing $C(l,\eta) = 0$ for all constraints $C$. When
dealing with constraints, we use the `$\approx$' sign to indicate an
equality, that holds only on this constraint surface and not
necessarily on the whole phase space. Thus by definition a constraint
function $C(l,\eta)$ satisfies
\begin{equation}
  \label{eq:notation:weakequality}
  C(l,\eta) \approx 0.
\end{equation}

\label{par:notation:graph}
The polygons that together make up a `tessellation', can be viewed as
a graph structure, consisting of faces (the polygons), edges (the
polygon edges) and vertices (the points where three or more edges
join). The sets of these faces, edges and vertices will be denoted by
$F$, $E$ and $V$ respectively. Single elements we denote by $f \in F$
and the number of elements by $\# F$. See
\fref{sec:dualgraph:dualgraph} for some more details.

\section*{A note on signs}

Some of the definitions used in this thesis differ by a minus sign
from referenced works or conventions; unfortunately it is not possible
to choose definitions to be simultaneously convenient, according to
conventions and equal to all referenced works. To keep things clear,
we give a (hopefully complete) list of differences.

The Poisson bracket has a relative minus sign with respect to the
definition in \cite{'tHooft:1993gz}; it is chosen to have the same
sign as the standard definition. Related to that is the definition of
the time evolution of a classical quantity (function on phase space) as
\begin{equation}
  \label{eq:notation:hamflow}
  \dot{f} = \poisson{f}{H},
\end{equation}
which is also according to normal conventions and opposite to the
definition used in \cite{'tHooft:1993gz} and related papers.

Boosts have the same sign as in \cite{'tHooft:1993gz}: contracting
edges have positive boosts. This looks a bit like a counter-intuitive
choice, but is necessary to get the right Hamiltonian flow for our
choice of signs
in~\fref{eq:notation:poisson},~\fref{eq:notation:hamflow} and the
Hamiltonian.

We use the space-like $(-,+,+)$ sign convention for the metric.

%%% Local Variables: 
%%% mode: latex
%%% TeX-master: "thesis"
%%% End: 

%% file: representation.tex
\chapter{Representing gravity}
\label{chap:representation}

Before we can start to think about the quantization of gravity in
arbitrary dimensions, we first have to choose a way to represent
gravity. We would like to be able to formulate gravity within the
framework of Lagrangian or Hamiltonian mechanics. Or in other words,
we want choose a set of variables that fully describe a state of the
system and then formulate the equations of motion on those states.

We must note however, that a classical ``state'' is normally defined
as a full description of the system for a given instance of time. But
in the context of general relativity ``a given instance of time'' is a
priori not an unambiguous statement to make: there is no unique time
coordinate, because coordinate diffeomorphisms are a symmetry of the
system.

In the next section we will introduce a way to split time and space in
general relativity, which then makes it possible to formulate general
relativity in a Lagrangian and then Hamiltonian formulation of
gravity.

\section{Foliation of space-time}
\label{sec:representation:foliation}

Before trying to split off time, we first have to give a mathematical
description of a space-time. There are two well known choices of
fundamental variables to describe a general space-time. We will use
the standard choice of taking the metric $g_{\mu\nu}(x^\sigma)$ as
field variable. There is however an alternative formulation, known as
`first order formalism' in which a set of orthonormal vectors
$e^a_\mu$ (called `dreibein', `vierbein' or in arbitrary dimensions
`vielbein') together with an internal connection is chosen as
fundamental variables. This description relates to the standard metric
description via
\begin{equation}
  \label{eq:representation:vielbein}
  \eta_{ab}\,e^a_\mu\,e^b_\nu = g_{\mu\nu}.
\end{equation}
In space-time dimension $n$, $\eta_{ab}$ is a diagonal matrix with
signature $(1,n\tm1)$, which expresses the fact that the $e^a_\mu$
are a local orthonormal set of vectors.

A space-time can mathematically be specified in a very general manner
as a differentiable manifold $M$ with a metric $g$ on it, often
denoted as a (pseudo-) Riemannian manifold $(M,g)$. In the following,
we will be dealing with a pseudo-Riemannian manifold of signature
(1,2).

\begin{wrapfigure}{r}{0cm}
%  \vspace{-0.3cm}
  \includegraphics{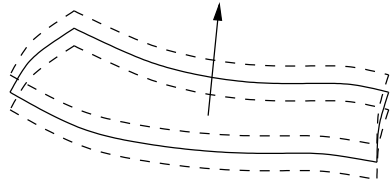}
  \caption{A foliation: the arrow indicates the direction of time.}
  \label{fig:foliation}
\end{wrapfigure}
We now want to formulate the theory in a Hamiltonian formalism, which
can later be used as a starting point for canonical quantization. For
this, we will choose an explicit time coordinate and introduce the
concept of a global foliation of space-time. This can be seen as a
continuous set of slices of space-time, where the chosen time
parameter labels the different slices; time has a fixed value on the
whole of each slice, see \fref{fig:foliation}. To this end, we
introduce a time function $T$ on $M$, associating with each point in
$M$ a unique time $t$. This function should be sufficiently
well-behaved such that its original $\Sigma_t = T^{-1}(t)$ for each
instant of time is a submanifold of $M$, which is spatial. It must
also be a monotonically increasing function along all timelike curves.

The process above amounts to a partial choice of
space-time\footnote{%
  One must be careful to make the distinction between coordinates that
  map space-time and coordinates that describe the state of the
  gravity system (which we will refer to as `configuration variables').} %
coordinates: we have chosen (only) a time coordinate, which slices the
manifold into a set of submanifolds of equal time. Note however that
this ``time'' is not necessarily the proper time for a local observer!
This complete set of submanifolds we now call a `time foliation' of
the manifold $M$. Note that there is no unique foliation: any time
function $T$ that satisfies some regularity properties will suffice
and give a different foliation.

Now that we have a time foliation, we want to pick a set of
configuration variables that for a given time $t$ specify the state of
the system, just like a function $x(t)$ would describe the state
(position) of a pointlike particle for a given time $t$. Heuristically
we can say that the system is the metric on the space-time manifold
$M$ and at a given instant of time $t$, it is given by a spatial slice
$\Sigma_t$ with the restricted metric. The configuration variables for
the complete manifold are given by $g_{\mu\nu}$ and for a fixed time
$t$ by restricting the metric to $\Sigma_t$ and only looking at the
space-space part of the metric: that is the part needed together with
$\Sigma_t$ to describe the state of space-time at time $t$. We also
need to specify the conjugate momenta to get a complete set of initial
conditions.

When we choose a complete set of space-time coordinates, we can make
this choice of configuration variables explicit by defining
\begin{equation}
  \label{eq:representation:submetric}
  g_{\mu\nu} = \left(\begin{array}{cc}
       - N^2 + N_k\,N^k    & \tilde{g}_{ij}\,N^i \\[10pt]
       \tilde{g}_{ij}\,N^j & \Big(\; \tilde{g}_{ij} \;\Big)
      \end{array}\right).
\end{equation}
The $N$, $N^i$ and $\tilde{g}_{ij}$ are functions of space-time that
can be read off from the explicit form of $g_{\mu\nu}$ in the chosen
space-time coordinates. We have chosen time to be a global timelike
coordinate function, so this implies that $g_{00} < 0$ and also that
the symmetric submatrix $\tilde{g}_{ij}$ for the other spatial
coordinates is non-degenerate and the entries $\tilde{g}_{ij}\,N^i$
thus well-defined.

The function $N$ is in the literature (see e.g.
\cite[p.~12]{Carlip:1998uc}) referred to as the `lapse function': it
specifies the rate that clocks tick with respect to the chosen time
coordinate. The functions $N^i$ are called the `shift functions' and
specify the displacement of spatial coordinates, when we make a small
displacement perpendicular to the surface $\Sigma_t$. The matrix
$\tilde{g}_{ij}$ can be viewed as a metric on the submanifold
$\Sigma_t$ and thus describes the state of the system (to be compared
to the position $x$ in a one-particle classical mechanical system).

Thus far, we have not yet made any specific choice of space-time
coordinates, except for the fact that we demanded a global time
coordinate. The only thing we have done is to make an explicit
separation of the time from the space coordinates to allow for a
treatment of the theory within a Lagrangian and Hamiltonian framework.
This decomposition is known as the Arnowitt-Deser-Misner (ADM)
formalism.

\section{Hamiltonian formulation}
\label{sec:representation:hamiltonian}

Using the ADM decomposition, we can write down the action of general
relativity and transform it to a Hamiltonian formulation. We will
here give an outline of this procedure, to give some insight in the
intricacies that will also show up in the polygon model. More details
can be found e.g. in reference \cite{Carlip:1998uc}.

The action of general relativity in three dimensions is given by
\begin{equation}
  \label{eq:representation:action}
  I = \frac{1}{16\,\pi\,G} \int \d^3\!x \; \sqrt{-\!\abs{g}} \,
         \left( R - 2\Lambda \right)
\end{equation}
in the absence of external matter or fields. We normalize
$16\,\pi\,G = 1$ and set the cosmological constant $\Lambda = 0$ from
here on.

Inserting the ADM decomposition of the metric, this action can be
rewritten as an integral over time $t$ of a Lagrangian $L$, with
\begin{equation}
  \label{eq:representation:adm-langrangian}
  L = \int_{\Sigma_t} \d^2\!x \; N \, \sqrt{\abs{\tilde{g}}} \,
  \left[ \tilde{R} + K_{ij}K^{ij} - \left(\tilde{g}^{ij}K_{ij}\right)^2 \right].
\end{equation}
Here $\tilde{R}$ is the Ricci scalar corresponding to $\tilde{g}$ and
$K_{ij}$ is the extrinsic curvature of $\Sigma_t$ embedded in $M$. In
principle there are also some boundary terms from partial integration,
but these cancel because the spatial slice is compact. The integrand
in \fref[plain]{eq:representation:adm-langrangian} we denote by
$\mathcal{L}$, the Lagrangian density.

Now we have a standard Lagrangian system and can make a transformation
to canonical momenta, which yields
\begin{align}
  \label{eq:representation:momenta}
  \pi^{ij} &= \pder{\mathcal{L}}{\left(\partial_t \tilde{g}_{ij}\right)}
            = \sqrt{\abs{\tilde{g}}}\,\Big[ K^{ij} -
              \tilde{g}^{ij}\left(\tilde{g}_{ab}K^{ab}\right) \Big],\\
  \label{eq:representation:adm-action}
  I &= \int \d t \int_{\Sigma_t} \d^2x \; \left(
    \pi^{ij}\,\partial_t\tilde{g}_{ij} - N\,\mathcal{H} - N_i\,\mathcal{H}^i \right)\\
\text{with}\qquad
  \label{eq:representation:ham-constr}
  \mathcal{H} &= \frac{1}{\sqrt{\abs{\tilde{g}}}}
  \Big[ \pi_{ij}\pi^{ij} - \left(\tilde{g}_{ij}\pi^{ij}\right)^2 \Big]
 -\sqrt{\abs{\tilde{g}}}\,\tilde{R},\\
  \label{eq:representation:mom-constr}
  \mathcal{H}^i &= -2 \, \tilde{\nabla}_j \, \pi^{ij}
\end{align}
The complete Hamiltonian, given by
\begin{equation}
  \label{eq:representation:hamiltonian}
  H = \int_{\Sigma_t} \d^2x \; \left( N \, \mathcal{H} + N_i \, \mathcal{H}^i \right),
\end{equation}
is now a linear combination of constraints, because the functions
$N, N^i$ do not have any equations of motion associated with them and
thus act as Lagrange multipliers for $\mathcal{H},\mathcal{H}^i$: the
variation of the action \fref{eq:representation:ham-constr} with
respect to $N, N^i$ yields $\mathcal{H} = 0,\, \mathcal{H}^i = 0$.

\section{Reduction of phase space}
\label{sec:representation:reduction}

We have found that in the Hamiltonian formulation we are left with
constraint functions. This means that the real dynamics of the
system takes place only in a subspace of the full phase space that we
had chosen initially.

We started with a configuration space $Q$ of all metrics
$\tilde{g}_{ij}$ on a spatial slice $\Sigma_t$ and the corresponding
phase space of canonically conjugate pairs
$(\tilde{g}_{ij},\pi^{kl})$. To get a description of only the true
physical degrees of freedom, we have to reduce this phase space. This
is done by first solving the constraint
functions~\fref[plain]{eq:representation:ham-constr}
and~\fref{eq:representation:mom-constr}.

In addition, the constraints above are first class constraints, so
they generate gauge transformations. The functions $\mathcal{H}^i$
generate coordinate transformations of the surface $\Sigma_t$, which
is a well known gauge freedom of general relativity. The function
$\mathcal{H}$ generates translations in time, relating coordinates on
different time slices $\Sigma$. This is in the strict sense not a true
gauge transformation, because it relates configuration variables at
different times. This fact is related to the problem of time in (the
quantization of) general relativity \cite{Isham:1992ms}.

Gauge transformations do not change the physical state of the system,
so as a second step, we have to fix this gauge freedom to get the
reduced phase space. The gauge freedom generated by the first class
constraints~\fref[plain]{eq:representation:ham-constr}
and~\fref{eq:representation:mom-constr} can now be removed by fixing
the Lagrange multiplier functions that were still arbitrary.

The explicit work of solving the constraints and fixing the gauge is
not as straightforward as it might seem from the formulation above:
there is no known way to solve this in $3\tp1$\ndash dimensions, but
in $2\tp1$\ndash dimensions there is
\cite{Moncrief:1989dx,Witten:1988hc}. We will not go into the details,
but with the use of Riemann surface theory one can consider the space
of metrics $\tilde{g}$ modulo diffeomorphisms and conformal factors,
which is a finite dimensional space known as the moduli space
$\mathcal{N}$ of $\Sigma$. This space has dimension $6(g\tm1)$ when
$\Sigma$ has genus $g>1$, two when $g = 1$ and zero when $g = 0$ (see
also \fref{fig:surfaces} for the concept of genus of a surface). The
physical phase space now has twice the dimension of this moduli space
$\mathcal{N}(\Sigma)$, because we still have to add the canonical
momenta, which span the same number of dimensions.

Thus in $2\tp1$\ndash dimensions, the physical phase space of general
relativity is finite dimensional, unlike the phase space we started
with: that was the space of all metrics $\tilde{g}$ and conjugate
momenta, which are fields over $\Sigma$ and thus infinite dimensional.
For this reason, general relativity in $2\tp1$\ndash dimensions is
sometimes called a `topological theory': the true degrees of freedom
do only arise from topological non-triviality of the surface $\Sigma$.
This again is a reason that the polygon model works: there are no
local degrees of freedom and locally space-time is flat. The
formulation of the polygon model depends on this fact as we will see
in \fref[plain]{chap:polygon}.

%%% Local Variables: 
%%% mode: latex
%%% TeX-master: "thesis"
%%% End: 

%% file: polygon.tex
\chapter{The polygon model}
\label{chap:polygon}

The polygon model is a model to describe $2\tp1$\ndash dimensional
space-times that can be foliated by a set of spacelike Cauchy
surfaces. This includes the possibility of spatially open and closed
universes \cite{'tHooft:1992vd}, although in the case of an open
universe one has to be careful about boundary conditions at spatial
infinity.

We will however only consider spatially closed universes. Thus if we
take a spatial slice of our complete space-time manifold $M$, then
this is a closed and bounded surface (so it has finite volume). More
specifically, we take a space-time of the form of a product:
\begin{equation}
  \label{eq:polygon:spacetime}
  M = \Sigma \times I.
\end{equation}
Here $I \subset \R$ is some interval of time, which may be bounded
from below and/or above, depending on whether there is a big bang or
big crunch respectively.

\begin{figure}
  \center
  \includegraphics{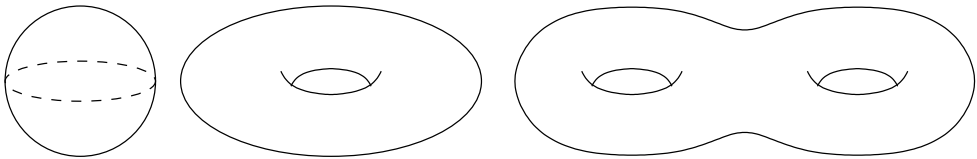}
  \caption{Surfaces of genus 0, 1 and 2.}
  \label{fig:surfaces}
\end{figure}

$\Sigma$ is the spatial part and thus should be a compact
two-dimensional surface. These surfaces are topologically completely
classified. The ones that are orientable are classified by their genus
$g$, which indicates the number of `holes' in them. The simplest
examples are a sphere, a torus and a `double torus', see
\fref{fig:surfaces}.

The model also allows for a finite number of pointlike, spinless
particles to be added to this space-time. We have not studied this in
detail and refer the reader to \cite{'tHooft:1993gz,Kadar:2004im} for
more information.

The simplest spatial surface of a sphere is only realizable with
particles. Proof for this fact can be found in
\cite[p.~57]{Carlip:1998uc} or \cite[p.~2912]{Moncrief:1989dx} and is
based on an analysis of the moduli space of metrics on the surface
$\Sigma$ modulo diffeomorphisms. For the sphere this moduli space
consists of one unique metric with curvature $1$, which does not yield
a solution to the vacuum Einstein equations. This can also be seen
from the fact that the Hamiltonian constraint can be written as in
\fref{eq:polygon:hamiltonian-constraint}:
\begin{equation*}
  H = 2\,\pi\,\chi = 4\,\pi\,(1-g).
\end{equation*}
The Hamiltonian equals minus the area of the dual graph as given
in~\fref{eq:dualgraph:area-triangle}. This area cannot be negative,
thus the Hamiltonian cannot be positive, which rules out the case
$g = 0$.

\section{Tessellation of the spatial surface}
\label{sec:polygon:tessellation}

We are considering general relativity in 3 dimensions. This we can
exploit to simplify the description of our space-time. In this case
the Riemann tensor is completely determined by the Ricci tensor: one
can show this e.g. by counting the degrees of freedom or explicitly
\cite[p.~3]{Carlip:1998uc}:
\begin{equation}
  \label{eq:polygon:riemann-ricci}
  R_{\mu\nu\rho\sigma} =
    g_{\mu\nu}   \,R_{\nu\sigma}
  + g_{\nu\sigma}\,R_{\mu\rho}
  - g_{\nu\rho}  \,R_{\mu\sigma}
  - g_{\mu\sigma}\,R_{\nu\rho}
  - \mfrac12 \left(g_{\mu\rho}  \,g_{\nu\sigma} -
                   g_{\mu\sigma}\,g_{\nu\rho}  \right)\,R.
\end{equation}
This means that in a region of space-time with no matter
($T^{\mu\nu} = 0$), the Einstein equations effectively reduce to
$R_{\mu\nu} = 0$ and determine that space-time is locally flat. Thus
everywhere we can locally choose a Minkowski coordinate system.

Given this space which is locally Minkowski, we can also easily choose
a spatial slice that is locally flat too. This will induce a local
coordinate system from the three-dimensional Minkowski space to our
two-dimensional spatial slice $\Sigma_t$. We cannot in general extend
this coordinate system to our whole spatial surface: our space is
compact, so when we try to extend our flat coordinate system from a
certain starting point, we will run into parts of space where we had
already defined coordinates. These do not necessarily have to match;
while we have extended our coordinates and run into a ``charted'' part
of space, we might have gone around a non-contractible loop in
$\Sigma$ which has a non-trivial holonomy. A simple example is a
toroidal spatial surface which expands uniformly in one of its
fundamental directions: when we walk along a loop in this direction
and return to our starting point, the coordinates fail to match by a
Lorentz boost of the speed of contraction or expansion.

\begin{figure}
  \vspace{-0.2cm}
  \centering
  \begin{minipage}[t]{.40\textwidth}
    \centering
    \includegraphics[width=6cm]{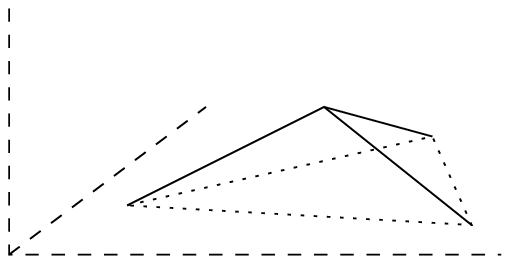}
    \caption{Curvature at a vertex of three joined $2$D surfaces in a
      flat $3$D background.}
    \label{fig:3d-curvature}
  \end{minipage}
  \hspace{1cm}
  \begin{minipage}[t]{.40\textwidth}
    \centering
    \includegraphics{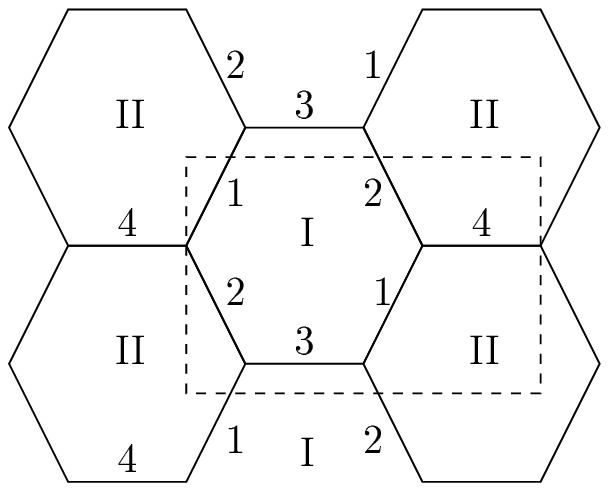}
    \caption{Tessellation of a torus by 2 polygons, labelled I and II.}
    \label{fig:tessellation}
  \end{minipage}
  \vspace{-0.3cm}
\end{figure}

We can however choose a coordinate system at some point(s) of our
spatial surface and extend it as far as possible. After doing this, we
will be left with several regions of flat space which, glued together
at their boundaries, form the complete spatial surface again. We will
call such a set of flat coordinate patches a `tessellation'. Notice
that at points where regions of space are glued together, the spatial
surface will in general not be flat. The underlying three-dimensional
Minkowski space-time is flat, but at points where three or more
spacelike slices join, the two-dimensional surface will in general
have a curvature of delta distribution type. See
\fref{fig:3d-curvature} for an example.

This is a tessellation of a spatial surface at one fixed value of the
time coordinate. The coordinate systems of the different patches
should therefore obey some boundary conditions, such that there is no
time difference when traversing a boundary and also time should run
equally fast on both sides. This implies that the boundary should move
perpendicular to its orientation in the coordinate frames on both
sides, which yields that a boundary must be a straight line and its
velocity be of same size and opposite direction in the neighboring
coordinate frames (see \cite[p.~1336]{'tHooft:1992vd} for a more
detailed discussion).

An example of a tessellation is given in \fref{fig:tessellation}: the
dashed lines show a fundamental domain of the torus (with the usual
identification of opposite sides) and two 6-sided polygons covering
this torus. The two polygons are labelled I and II and the different
edges are labelled 1 to 4. Notice that these polygons have a number of
edges in common with each other and also with themselves.

Thus our tessellation consists of a number of polygon coordinate
patches with sides that have to be identified. The sides of these
polygons have well-defined lengths within the coordinate frames of the
respective polygons. These should however match the length of the edge
as seen in the coordinate frame on the opposite side of the boundary
(because the metric should be continuous). Therefore these boundary
edges have a unique, well-defined length $l$.

Besides that, we can assign a boost parameter to each edge. This boost
is $\eta = \tanh^{-1}(v)$, where $v$ is the speed of the edge in its
neighboring coordinate frames. The speed and boost have a positive
sign when the edge moves inward, thus $v$ is the speed at which the
edge contracts the polygon. This speed is uniquely defined, because an
edge must expand or contract with the same speed in the coordinate
frames on both its sides. The Lorentz transformation relating two
neighboring coordinate frames is then given by a boost of $2\,\eta$:
first a boost from one frame to the rest frame of the edge and then
again a boost from the edge rest frame to the other coordinate frame.

\section{Vertex relations}
\label{sec:polygon:vertexrelations}

We have made a tessellation of our spatial slice with polygons. These
polygons exactly cover this spatial slice, so boundaries (`edges') of
these polygons have to be identified where they are ``glued together''.

Besides that we also have `vertices': points where three or more edges
join. As was stated earlier, space-time is locally flat. This should
also hold at vertices. To express the curvature located at a vertex,
we can write down the holonomy around that vertex: the change a vector
undergoes when it is being parallel transported around that vertex.

We can explicitly write down the complete holonomy of going around the
vertex using the Lorentz matrices~\fref{eq:notation:lorentzmatrices}.
Going around in anti-clockwise orientation, we label the polygons at
the vertex $i = 1 \ldots n$ and define $\alpha_i$ as the angle that
polygon $i$ makes at the vertex and $2\eta_i$ the boost between
polygon $i$ and polygon $i+1 \mod n$. Now going around means
successively making a rotation $L_r(\alpha_i)$ and a boost
$L_x(2\eta_i)$ for each polygon.\footnote{%
  Remember that the transformation between two neighboring polygons
  goes with twice the boost of the edge.  Furthermore the precise
  choice of signs and $L_x$ or $L_y$ depends on the explicit choice of
  coordinates in the polygons and orientation of the holonomy, but
  this does not affect the resulting equations. This can be seen from
  the fact that the system has space and time mirror symmetry.} %
This should be equal to the identity operation, as space-time must be
flat at the vertex point, thus we have
\begin{equation}
  \label{eq:polygon:vertexholonomy}
  \prod_{i=1}^n L_r(\alpha_i)\,L_x(2\eta_i) = \mathlarger{\mathbbm{1}}.
\end{equation}
We demand that at a vertex always exactly 3 polygons join. This is
always possible by choosing a suitable tessellation and will allow us
to express the angles in terms of the boosts. For later convenience,
we now choose the indices of boosts and angles in a more symmetric
fashion as in \fref{fig:vertex-params}. When we rewrite
\fref{eq:polygon:vertexholonomy} to a form with 3 matrices on the left
and the right side and then explicitly work out the equations, we get:
\begin{subequations}\label{eq:polygon:vertexrelations}
\begin{align}
  \label{eq:polygon:vertexrelations-1}
         & s_1 : s_2 : s_3 = \sigma_1 : \sigma_2 : \sigma_3,\\
  \label{eq:polygon:vertexrelations-2}
0        &= s_1 \, c_2 + \gamma_2 \, s_3 + c_1 \, s_2 \, \gamma_3,\\
  \label{eq:polygon:vertexrelations-3}
c_1      &= c_2 \, c_3 - \gamma_1 \, s_2 \, s_3,\\
  \label{eq:polygon:vertexrelations-4}
\gamma_1 &= \gamma_2 \, \gamma_3 + c_1 \, \sigma_2 \, \sigma_3
\end{align}
\end{subequations}
and all cyclic permutations of the indices. Here we used the notation
as in \fref{eq:notation:boost-shorthand}. As in the literature, we
will refer to these as the `vertex relations'. These equations are not
independent \cite{Hollmann:1999jd} nor have we given the same set as
originally given by 't~Hooft in \cite{'tHooft:1992vd}, but they are
complete in the sense that they are equivalent to the full set of
equations determined by \fref[plain]{eq:polygon:vertexholonomy}.

These equations now give us expressions for the angles of polygons in
terms of boost parameters or vice versa. To determine the angles one
can for example rewrite
equation~\fref[plain]{eq:polygon:vertexrelations-4} and take the
inverse cosine. This does not yet determine the angle uniquely, but
with the help of equation~\fref[plain]{eq:polygon:vertexrelations-1}
it does. There is the extra fact that at most one of the three angles
at a vertex can be larger than $\pi$, because if there would be two
angles larger than $\pi$, there would be points that show up in
the coordinate systems of both polygons \cite{'tHooft:1993gk}. This
then determines the sign of the sines of the angles at each vertex.

\section{Constraints}
\label{sec:polygon:constraints}

We now have a representation of a 2D spatial slice of the 2+1D
space-time manifold $M$ in terms of one or more polygons, glued
together at their edges. A configuration of such a spatial slice is
described in terms of the lengths $l_i$ and boost parameters $\eta_i$
of all edges together with the graph structure of all these edges.

The $l_i, \eta_i$ however cannot take all possible values in
$\R^{2N}$. Besides the fact that all lengths have to be positive
($l_i \ge 0$), there are some other, more complicated constraints.

Firstly the boosts at each vertex have to obey a triangle inequality.
For each triple of boosts at one vertex and each permutation thereof,
the following must hold:
\begin{equation}
  \label{eq:polygon:triangle}
  \abs{\eta_a} + \abs{\eta_b} \ge \abs{\eta_c}
\end{equation}
This can be deduced from vertex
relation~\fref{eq:polygon:vertexrelations-4}. A more geometrical
argument is as follows: when going around the vertex, we are Lorentz
boosted three times, but we should return to our original rest frame.
The resulting boost of two boosts is at most the sum of the boosts,
exactly when they are collinear. So the sum of lengths of each two
boosts must be greater than the third to be able to close the
holonomy. This is analogous to the case of Galilean transformations
when we replace the boosts by normal velocities.

Secondly there is a set of constraints for each polygon. Geometrically
these are easy to formulate: each polygon must close. This means that
the angles and edge lengths must be such that when going around the
border of a polygon and keeping track of the relative coordinates,
that one must end up exactly where one started and the sum of the
angles must be $2\,\pi$.

As the angles can be expressed in terms of the boosts, we can
explicitly express these constraints in terms of the model variables
$l_i, \eta_i$. For a fixed polygon, let us number the lengths, boosts
and angles according to \fref{fig:polygon-params}.

We first define the angle between edge $j$ and the horizontal (edge
$N$) by
\begin{equation}
  \label{eq:polygon:theta}
  \theta_j = \sum_{i=1}^j \pi - \alpha_i.
\end{equation}
Using this we can write down the angular and closure constraints as
\begin{subequations}\label{eq:polygon:constraints}
\begin{align}
  \label{eq:polygon:constraint-angle}
  C_\theta &= \theta_N - 2\,\pi = \sum_{j=1}^N (\pi - \alpha_j) - 2\,\pi \approx 0,\\
  \label{eq:polygon:constraint-complex}
  C_z &= \sum_{j=1}^N l_j\,\exp( i \, \theta_j ) \approx 0.
\end{align}
\end{subequations}
The closure constraint $C_z$ is written using an implicit complex
coordinate system in the polygon. This way of writing is obviously
equivalent to formulating two real-valued constraints for the $x$ and
$y$ direction. We will refer to the second constraint as the `complex
constraint' as in \cite{Welling:1996hg}.

\section{Dynamics}
\label{sec:polygon:dynamics}

From the definition of our length and boost parameters, we can find
the evolution of these parameters. The boosts do not change in time,
except when transitions take place. The lengths however do change. A
geometrical analysis shows that the length change of an edge gets a
contribution from the vertices at both its sides equal to
\begin{equation}
  \label{eq:polygon:dlength}
  \der{l}{t}
= -\frac{\tanh(\eta_1)\cos(\alpha_3) + \tanh(\eta_2)}{\sin(\alpha_3)}
= -\frac{v_1\,c_3 + v_2}{s_3},
\end{equation}
where index 1 numbers the edge under consideration and indices 2 and 3
may be interchanged (there is mirror symmetry).

We can also look at the evolution of the system in the Hamiltonian
formulation. The Hamiltonian is given by the total two-dimensional
curvature of the spatial slice. Curvature is only present at vertices
and is given there by the deficit angle, thus
\begin{equation}
  \label{eq:polygon:hamiltonian}
  H = \sum_{v \in V}\left( 2\,\pi - \sum_{i=1}^3 \alpha_{v,i} \right).
\end{equation}
We can see that this Hamiltonian is a constraint: on the one hand, it
is the integral over the scalar curvature, which is only dependent on
the Euler characteristic $\chi$ (see
formula~\fref{eq:dualgraph:eulerchar}) by the Gauss-Bonnet theorem:
\begin{equation}
  \label{eq:polygon:hamiltonian-gaussbonnet}
  H = \int_\Sigma \d^2\!x \; \sqrt{\tilde{g}}\,\tilde{R}
    = 2\,\pi\,\chi = 4\,\pi\,(1-g).
\end{equation}
On the other hand we can see that it is equal to the sum of all
angular constraints, using the Euler characteristic of the graph:
\begin{align*}
0 &= \sum_{f \in F} \left( C^f_\theta - 2\,\pi \right)\\
  &= - \sum_i \alpha_i + \left(3\,\#V - 2\,\#F \right) \pi \\
  &= \sum_{v \in V} \left(2\,\pi - \sum_{i=1}^3 \alpha_{v,i}\right)
    +\left(\#V - 2\,\#F\right) \pi\\
  &= H - 2\,\pi\,\chi.
\label{eq:polygon:hamiltonian-constraint}\eqnumber
\end{align*}
In the last step we made use of the trivalence of the vertices by
plugging in the relation $3\,\#V = 2\,\#E$. See
page~\pageref{par:notation:graph} for notation.

The Hamiltonian now generates evolution of the system by
\begin{equation}
  \label{eq:polygon:flow}
  \begin{split}
  \der{l_i   }{t} &= \poisson{l_i   }{H},\\[2pt]
  \der{\eta_i}{t} &= \poisson{\eta_i}{H}.
  \end{split}
\end{equation}
To explicitly calculate this, we need a symplectic structure on phase
space. A symplectic structure defines Poisson brackets and it is
uniquely determined by the Poisson brackets between the fundamental
variables.  In our case it turns out that the evolution of the boosts
and lengths exactly matches the evolution generated by the Hamiltonian
if we choose the symplectic structure to be
\fref[plain]{eq:notation:poisson}:
\begin{equation*}
  \poisson{l_i}{\eta_j} = \poissonleta.
\end{equation*}

The Hamiltonian time evolution of the boost parameters can immediately
be seen to match that of the geometrical picture: in both cases it is
zero. For the geometrical picture it was chosen this way by letting
time run equally fast everywhere in each polygon. For the Hamiltonian
evolution, we see that it follows from the fact that $H$ only depends
on the boost parameters.

\clearpage
\section{Transitions}
\label{sec:polygon:transitions}

Given the results of the previous section, the evolution of the system
looks simple: the boosts and angles are fixed and the lengths change
linearly in time. It turns out that things are not that simple: there
is a complicating factor coming from the constraint that the model
parameters must describe a true tessellation. This means that edge
lengths must be positive and polygons must be true,
non-self-intersecting polygons.

As a side remark, it must be noted that strictly speaking, polygons
may be self-intersecting, but only in such a way that they can be
embedded in a non-self-intersecting way on some two-dimensional
surface, see \cite{'tHooft:1993nj}. This can be formulated as follows:
if a polygon intersects itself then this will generate a transition if
and only if a new triangle is formed and the winding number of points
that will lie in the triangle decreases with respect to the oriented
boundary of the polygon. If the winding number of these points
decreases, then these points are on the outside of both parts of the
polygon boundary that cross and thus would start to show up in two
different places in polygon coordinate charts.

\begin{figure}[b]
  \centering
  \begin{minipage}[t]{.30\textwidth}
    \centering
    \includegraphics{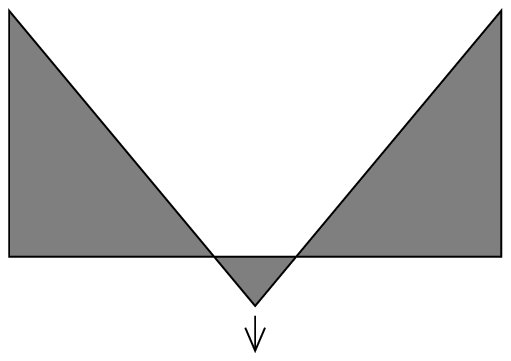}
    \caption{A truly self-intersecting polygon.}
    \label{fig:polygon-intersect}
  \end{minipage}
  \hspace{2cm}
  \begin{minipage}[t]{.35\textwidth}
    \centering
    \includegraphics{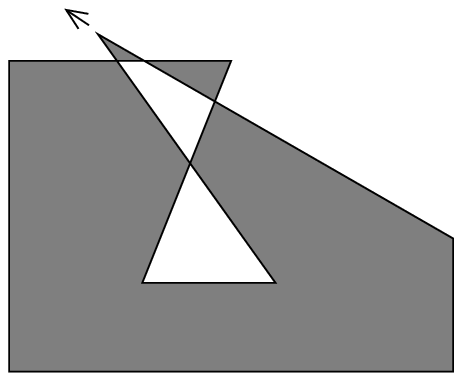}
    \caption{A superficially self-intersecting polygon.}
    \label{fig:polygon-non-intersect}
  \end{minipage}
  \vspace{0.5cm}
\end{figure}
\Fref[plains]{fig:polygon-intersect}~and~\fref[plain*]{fig:polygon-non-intersect}
show two polygons that both seem to self-intersect. In the first a
transition will split the polygon into two parts when the vertex hits
the horizontal edge. The winding number of points in the new triangle
decreases from $0$ to $-1$. In the second figure, no transition will
take place when the vertex crosses the edges as indicated.

\begin{figure}
  \center
  \includegraphics{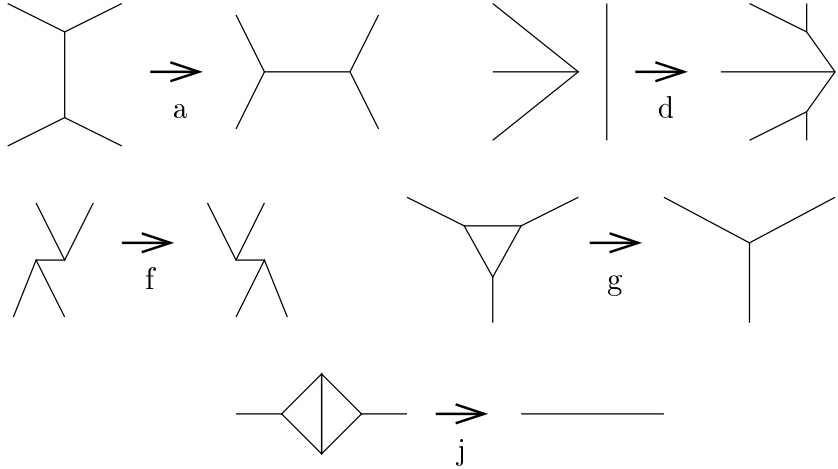}
  \caption{All possible polygon transitions without particles,
    labelled as in \cite{'tHooft:1993nj}.}
  \label{fig:transitions}
\end{figure}
Returning from this digression, we now have essentially two
transitions that can happen: an edge collapsing to zero length (the
top left diagram in \fref{fig:transitions}) and a vertex hitting an
edge (top right). There are some special cases of these two
transitions. If an edge is part of a triangular polygon, then this
whole polygon will disappear when the edge length goes to zero. This
is due to the fact that boosts and angles are constant and thus a
triangle will scale to zero size when one of its edges does. This can
also happen with two neighboring triangles (but not with three, as
they would together close the spatial slice, but violate initial
conditions). See the middle right and bottom diagrams of
\fref[plain]{fig:transitions}. Furthermore we have a special case when
an edge shrinks to zero, but the angles are such that it starts
growing again, but with angles changed by $\pi$ (called a `grazing
transition' in the literature); see the middle left diagram of
\fref[plain]{fig:transitions}.

There are a few more possible transitions when one allows particles in
the model. For the full set of transitions including those with
particles we refer to \cite{'tHooft:1993nj}.

Now if we have a set of initial conditions that satisfy the
constraints, we can calculate the time evolution. This is in first
instance given by only the length change of the edges. But after a
certain amount of time, the polygon configuration may have changed in
such a way that a transition as depicted in \fref{fig:transitions} is
about to take place. This will not only change the graph structure as
indicated, but also the model parameters will change. One can find
these by demanding that the parameters after the transition should be
consistent with those before and that they should obey the vertex
relations: space-time must stay flat.

A potential problem now arises: when a transition takes place, we
switch to a different graph structure with a different set of edges
and thus a different phase space to describe the system. The number of
edges does not need to be the same, so the dimension of the phase
space can change during a transition. The dimension of the physical
phase space does not change (as expected). This we can see from
counting degrees of freedom and constraints. The Euler characteristic
(see \fref{sec:dualgraph:dualgraph}) is
\begin{equation}
  \label{eq:polygon:eulerchar}
  \chi(\Sigma) = \#V - \#E + \#F
\end{equation}
and is constant for a given topology of $\Sigma$. From the trivalence
of the vertices, we obtain $3\,\#V = 2\,\#E$. This together gives us
the relation
\begin{equation}
  \label{eq:polygon:nedges-nfaces}
  \#F - 3\,\#E = \text{constant},
\end{equation}
which does not depend on the specific structure of the graph. Now two
degrees of freedom are associated with each edge. Each polygon on the
other hand introduces three constraints. And as we have seen in
\fref{sec:representation:reduction}, each constraint generates a
complementary gauge degree of freedom. So each polygon amounts to a
total of six unphysical degrees of freedom in the full phase space,
which exactly cancels the extra degrees of freedom added by the edges.

\section{Constrained dynamics}

With the addition of transitions, we now have a complete description
of the (classical) dynamics of the system: given a set of initial
conditions, we can uniquely determine the evolution of the system.
This system must not only be uniquely defined, but also be
well-defined. For this we have to check that all constraints
(equalities and inequalities) are preserved under the evolution of the
system.

\subsection{Equality constraints}

We want to check that time evolution preserves the constraints, which
amounts to checking that $\poisson{C}{H} \approx 0$ for all
constraints.  This follows trivially for the angular constraints,
because they do not depend on the lengths. The calculation for the
complex constraints is not straightforward, but also gives
$\poisson{C_z}{H} \approx 0$~: see
\fref[plain]{chap:constraintalgebra} for the details.

Besides this, one must also check that the constraints are preserved
during transitions. This turns out to be relatively simple for the
complex constraint: during transitions of the type where an edge
length reduces to zero (transitions a,~f,~g,~j in
\fref{fig:transitions}), only edges with length zero are removed and
created. Thus these do not affect the complex constraints. Also in the
case of transition type d, the complex constraint is conserved: the
new lengths of the split edges are by construction chosen to match the
complex constraint.

\begin{figure}
  \center
  \includegraphics{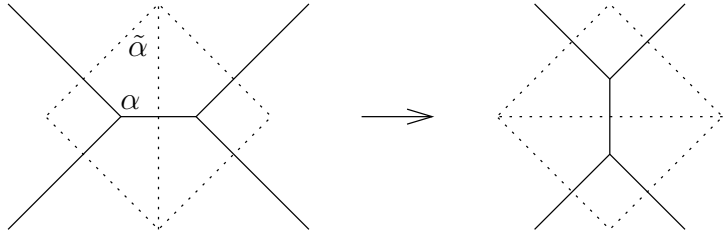}
  \caption{An edge transition and the dual graph (dotted lines).}
  \label{fig:trans-edge-dual}
\end{figure}

The angular constraints are also preserved under transitions. The best
way to understand this is by looking at the dual graph (see
appendix~\fref[plain*]{chap:dualgraph} and
\cite{Hollmann:1999jd,Kadar:2003ie}). This graph is constructed by
mapping vertices into faces and vice versa. Edges are mapped to
themselves and to each edge in the dual graph we associate an
(oriented) length $\tilde l = 2\,\eta$, where $\eta$ was the boost of
the edge in the original graph. To each angle $\alpha$ we associate an
angle $\tilde\alpha = \pi - \alpha$ in the dual graph as shown in
\fref{fig:trans-edge-dual}.

Now a transition will also induce a corresponding transition in the
dual graph. As explained in appendix~\fref[plain*]{chap:dualgraph},
the dual graph (with the mapping of edge lengths and angles) can be
embedded in the hyperbolic plane. From the transitions in the dual
graph, it can now be shown that the angular constraints are preserved

Making an edge transition of type a (\fref{fig:transitions}) is now a
matter of removing an edge in the dual graph and replacing it by the
opposite edge within the quadrilateral just created, as in
\fref{fig:trans-edge-dual}. This dual graph is embedded in hyperbolic
space, where every point has a full $2\,\pi$ neighborhood (as in flat
space). Each neighborhood of a vertex in the dual graph is also fully
covered before the transition, as can be deduced from the angular
constraint of the polygon corresponding to that vertex.

The other transitions can be dealt with in a similar fashion: the
angular constraints are encoded in the dual graph as vertices
containing a $2\,\pi$ neighborhood and making the corresponding
transition in the dual graph does not change this, thus the transition
should preserve the angular constraint. We have not checked this
explicitly and there might arise some problems, when mixed vertices
are involved, see \fref{sec:dualgraph:embedding}.

\subsection{Inequality constraints}

In \fref{sec:polygon:constraints} we saw that there are also some
inequalities to be satisfied, namely, lengths must be positive and the
boost parameters must obey triangle
inequality~\fref[plain]{eq:polygon:triangle}. This raises some
problems when trying to find a quantized version of this model.

In a standard Hamiltonian formulation, we have a configuration space
$Q$ and a state of the system is given as a point in the cotangent
space $T^\star Q$. Normally $Q$ is chosen to consist of the positions
(lengths in our case) and then elements in the vector space
$T_q^\star Q$ are the momenta (the boosts in our case). We now face
the problem that the Hamiltonian is only defined for those boosts
that satisfy the triangle inequality, and thus $H$ is not a function
on the whole phase space $T^\star Q$.

\begin{figure}
  \centering
  \begin{minipage}[b]{.40\textwidth}
    \center
    \includegraphics{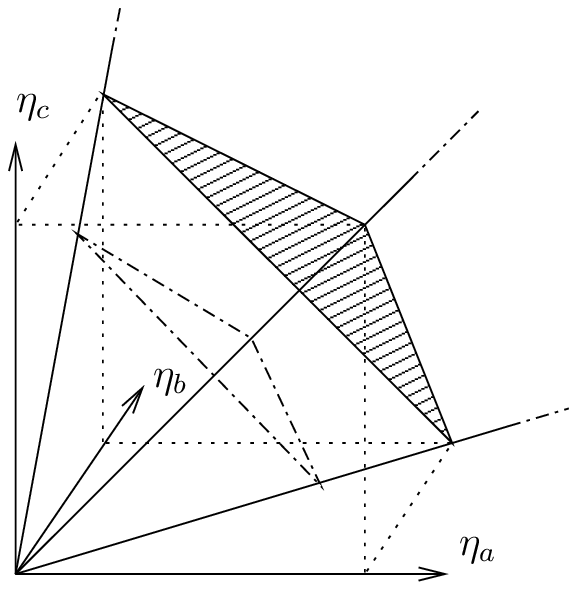}
    \caption{The restriction of boosts by the triangle inequalities.}
    \label{fig:boosttriangle}
  \end{minipage}
  \hspace{1.5cm}
  \begin{minipage}[b]{.35\textwidth}
    \center
    \includegraphics{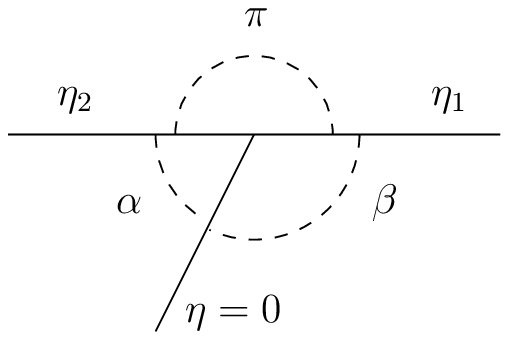}
    \caption{Parameters at a vertex with one zero boost.}
    \label{fig:vertexboost0}
  \end{minipage}
\end{figure}

In \fref{fig:boosttriangle}, a picture is given of the triples of
boosts $(\eta_a,\eta_b,\eta_c)$ at one vertex that satisfy these
inequalities. These are the triples within the triangles like the one
shaded in the figure, stretching out radially. Note that the picture
does only display the triples in the first octant of $\R^3$ (all
boosts positive), so the full picture should be mirrored in three
planes. Moreover, it only shows the constraints from one vertex; each
edge is connected to two vertices, making the complete set of
inequalities more complicated.

Special care has to be taken at the boundary of the inequalities, when
they are exactly satisfied. Here we can distinguish two cases
according to whether one of the three boosts $\eta_a, \eta_b, \eta_c$
is zero or not. If none of the three is zero, the angle opposite to
the largest boost is zero, which does not correspond to an acceptable
tessellation.

If one of the boosts is zero, the other two must be equal (also in
sign) and a degree of freedom in the boosts is replaced by a degree of
freedom in the angles at that vertex. This can easily be seen from the
geometrical picture (see \fref{fig:vertexboost0}): if one edge boost
is zero, the polygons separated by that edge can be joined together as
the Lorentz transformation between those polygons is trivial. Then the
two other edges with boosts $\eta_1, \eta_2$ become one edge and thus
should have the same boosts $\eta_1 = \eta_2$ and angle $\pi$ between
them. In this case, the angle $\alpha$ is not determined by the vertex
relations and can be chosen freely. Given $\alpha$, we have
$\beta = \pi - \alpha$. If we also take $\eta_1 = 0$, the triangle
inequality implies that also $\eta_2 = 0$ and the three angles are
free but should add up to $2\,\pi$.

This change from boosts to angles to parametrize the degrees of
freedom in phase space seems to indicate that the $l_i,\eta_j$
coordinates cannot cover the full phase space. The subspace not
covered seems to be lower-dimensional though.

\section{Simulation}

Lastly, it must be mentioned that this polygon model has as one of its
virtues that it is well suited for numerical simulation
\cite{'tHooft:1993gz}. The model has an explicit description of the
evolution of the system, which is finite dimensional and not in terms
of differential or integral equations. This allows one to ``exactly''
simulate it on a computer: the exactness is only bound by the finite
precision used in the simulation.

To simulate a space-time, one must first find a set of parameters that
satisfy the constraints. This is in general very hard, but an explicit
(but intricate) method for genus $g > 1$ has been given by Kadar and
Loll in \cite{Kadar:2003ie} using the dual-graph representation (see
appendix~\fref[plain*]{chap:dualgraph}). A more simple approach is to
start with a completely symmetric one polygon tessellation (OPT) and
perturb the parameters slightly. One can then pick one boost and two
lengths and solve the angular and complex constraints for those. This
approach does not yield initial conditions for the full phase space
however.

Now, one can simulate the polygon model by calculating the length
changes $\dot{l}_i$ from \fref{eq:polygon:dlength} and let time run.
If this length change will result in one of the transitions in
\fref[plain]{fig:transitions}, one has to ``stop'' the simulation and
perform the transition, by changing the graph structure and
calculating the new parameters using the vertex relations
\fref{eq:polygon:vertexrelations}. This can then be repeated until we
either run into a big crunch, or there are no transitions taking place
anymore and space expands to infinity. One can also reverse time by
changing sign of all boosts, to simulate the past of a given
configuration.

We mentioned that this simulation is exact up to numerical precision.
Unfortunately, if the universe collapses towards a big crunch, the
boosts tend to grow exponentially under transitions and precision is
lost after only a couple of transitions. This inaccuracy cannot be
prevented easily as the boosts are calculated using hyperbolic sines
and cosines, thus making intermediate expressions roughly exponentials
of the boosts and absolute errors blow up equally fast.

We have implemented a simulation program too, to gain a better
understanding of the details of the model and with the hope of
revealing some interesting aspects of $2\tp1$\ndash dimensional
gravity for genus $g > 1$. This last goal has showed to be fairly
difficult: the interpretation and visualization of simulation data is
hard and especially the extraction of physical information turned out
to be difficult. The observations that we made in this simulation as
mentioned above, do coincide with those already made by 't~Hooft in
\cite{'tHooft:1993gz}.

%%% Local Variables: 
%%% mode: latex
%%% TeX-master: "thesis"
%%% End: 

%% file: quantization.tex
\chapter{Quantization}
\label{chap:quantization}

The polygon model for $2\tp1$\ndash dimensional gravity is a
relatively simple model, with only a finite number of degrees of
freedom. Furthermore we have been able to explicitly formulate it in a
Hamiltonian formalism.  This raises the question whether it is
possible to quantize this model. Even though a full four-dimensional
quantum gravity theory would still be far away, it might give
interesting indications about qualitative aspects of quantization of
general relativity.

One such thing is the question whether space and/or time will become
discretized. Already for this model there are different predictions:
't Hooft argues in \cite{'tHooft:1993nj} that the imaginary part of
(analytically continued) lengths will be quantized and that time will
be discretized. This means that the time evolution is well-defined for
discretized time steps only. Waelbroeck on the other hand argues in
\cite{Waelbroeck:1996sg} that this time discretization can be lifted
by choosing an internal time: the Hamiltonian constraint is solved for
one of the boosts and this boost, expressed in terms of the other
boosts, will then take the role of the Hamiltonian and its conjugate
edge length will take the role of time.

As already argued in the papers cited above, the quantized version of
this theory might very well depend on the way we quantize. We have a
set of constraints, which we can choose to first solve and only then
quantize the reduced phase space. On the other hand, we can also first
quantize and afterwards impose these constraints as operators in the
quantum theory.

These predictions about a quantum theory of the polygon model are
heavily dependent on the way one quantizes the theory. We will be
concerned with finding a canonical quantization only and not consider
other quantization methods like path integral quantization: canonical
quantization is already a non-unique quantization scheme, that is also
not guaranteed to work. Therefore we will first investigate the a
priori question of whether a consistent canonical quantization exists, 
before further investigating the predictions of a quantized model.

\section{General procedure}
\label{sec:quantization:general}

Here, we will describe what it means in a general setting to
canonically quantize a classical system.

We start with a classical system with a configuration space $Q$, which
is locally isomorphic to $\R^n$, or in other words, which is an
$n$\ndash dimensional smooth manifold. This also naturally defines a
phase space as the cotangent space $P = T^\star Q$, in which we can
identify a pair of position and momentum coordinates $(x,p)$ as the
base point $x \in Q$ and $p \in T_x^\star Q$ as a linear form on the
tangent (velocity) vectors at $x$.

On this phase space in turn is defined a Poisson bracket structure,
which maps pairs of functions on $P$ to new functions on $P$, defined
in local coordinates $(x,p)$ as
\begin{equation}\label{eq:quantization:poisson}
  \poisson{f}{g} = \sum_{i=1}^n
  \left( \pder{f}{x_i}\pder{g}{p_i} - \pder{f}{p_i}\pder{g}{x_i} \right).
\end{equation}
A Hamiltonian, defined as a function on the phase space, then describes
the classical evolution of the system by the flow of the Hamiltonian
vector field: $\chi_H = \poisson{.}{H}$.

Now quantization amounts to the following. We want to map classical
observables to quantum operators. These operators must act on wave
functions in a vector space with inner product, that is, a Hilbert
space. In the following, we will construct the picture bottom up; this
is contrary to the way one would construct a quantum theory in
practice, but allows us to specify the mathematics involved more
rigorously.

We consider a Hilbert space $S$ of wave functions mapping from the
configuration space $Q$ to the complex numbers. The inner product on
this Hilbert space defines normalization of wave functions. Normally
this Hilbert function space is chosen to be the square integrable
functions $L^2(Q;\C)$.

Within this Hilbert space, we consider a finite dimensional Lie group
$G$ within the set of unitary operators $U: S \mapsto S$. Such a Lie
group $G$ has a Lie algebra $\mathcal{G}$ associated with it, which
carries the standard anti-symmetric multiplication structure, called
the Lie brackets, such that for all $f,g,h \in \mathcal{G}$:
\begin{equation}\label{eq:quantization:liealg}
  \begin{split}
    \lie{f}{g}   &= -\lie{g}{f}\\
    \lie{f+g}{h} &=  \lie{f}{h} + \lie{g}{h}\\
    0 &= \lie{f}{\lie{g}{h}} + \lie{g}{\lie{h}{f}} + \lie{h}{\lie{f}{g}}
  \end{split}
\end{equation}

We are then looking for a set of classical observables, that is,
functions on the phase space $P$, a Lie group and its corresponding
algebra $\mathcal{G}$ and a mapping from these classical observables
$O_i$ to Lie algebra elements $\mathcal{O}_i \in \mathcal{G}$, for
which the Poisson brackets are mapped to $-i$ times\footnote{%
  This depends on conventions: physicists tend to add the $-i$, with
  the effect of the operators becoming Hermitian, thus having real
  eigenvalues. The operators in the Lie algebra however have to be
  anti-Hermitian to give unitary operators when exponentiated.} %
the Lie brackets. In this mapping of the classical operator Poisson
algebra to a Lie algebra, one normally also includes a `quantum
deformation'. This means that these relations are modified by adding a
factor like $\hbar$ to introduce a quantum scale.

This process of canonical quantization is by no means a straightforward
procedure: the choice of Hilbert space, classical operators and
mapping to quantum operators is not prescribed. One must just try
and see whether the choice made gives a consistent scheme.

In ordinary quantum mechanics on $\R^n$, one promotes $x$ and $p$ to
operators, which yields the well-known commutation relations
\begin{equation}
  \label{eq:quantization:xp-commutator}
  \left[ \hat{x}_i , \hat{p}_j \right]
  = i\,\hbar\,\widehat{\poisson{x_i}{p_j}}
  = i\,\hbar\,\delta_{i,j}\,\hat{\mathbbm{1}}.
\end{equation}
Here we see that a quantum deformation of a factor $\hbar$ has been
added.

This is then combined with the Hilbert space $S = L^2(\R^n;\C)$ of
square-integrable wave functions over the positions $x_i \in \R^n$.
The operators $\hat{x}_i$ and $\hat{p}_j$ are realized on this Hilbert
space in the position representation as the self-adjoint operators
\begin{equation}
  \label{eq:quantization:xp-operators}
  \begin{split}
    \hat{x}_i \; \phi(x) &= x_i\,\phi(x),\\
    \hat{p}_j \; \phi(x) &= -i\,\hbar\,\pder{\phi(x)}{x_j}.
 \end{split}
\end{equation}

\section{Constraints in quantization}
\label{sec:quantization:constraints}

In the outline above, we have not yet talked about constraints. These
do not show up in the canonical quantization of a particle: there all
degrees of freedom are truly physical degrees of freedom. In the
polygon model we have to accommodate constraints. This can be done in
essentially two ways: we can either implement those at the classical
level and then quantize, or we can first quantize the theory and only
then implement the constraints quantum mechanically.

Implementing them at the classical level means that we solve the
constraints classically by constructing a fully reduced phase space.
Thus we have to reparametrize the phase space coordinates of the
system in such a way that the constraints are fulfilled by
construction.

We must be careful in solving the constraints: the theory was
formulated as a generally covariant theory, which means that if we
solve all constraints, we are left with a so called `frozen time'
picture. The Hamiltonian itself was a constraint in this formulation,
so time evolution is a (semi) gauge transformation. This means that if
we naively solve the Hamiltonian constraint, then in the fully reduced
phase space, the Hamiltonian constraint is just a constant, so the
Hamiltonian equations of motion become trivial. The state space of the
system is then exactly parametrized by a complete set of constants of
motion.

This problem of a frozen time picture can be overcome by breaking
general covariance. We can introduce a time again (often referred to
as `internal time') by choosing a classical observable that is
monotonic in the original time parameter. The simplest choice for such
a parameter is one of the edge lengths. Let us label this length and
its conjugate boost by $(l_0,\eta_0)$. The action
\begin{equation}
  \label{eq:quantization:action-full}
  S[(l_0,l_i,\eta_0,\eta_i)(t)] =
  \int \d t \left[
    2\,\eta_0\,\dot{l}_0 + \sum_i 2\,\eta_i\,\dot{l}_i
  - u\,H
  - \sum_{f \in F} \left( v_f\,C^f_\theta + w_f\,C^f_z \right)
  \right]
\end{equation}
then yields the correct equations of motion. We see that the
Hamiltonian and all constraints have Lagrange multipliers
($u,v_f \in \R ,\, w_f \in \C$) associated with them; the equations of
motion for these multipliers result in the constraint equations, cf.
\fref{eq:representation:hamiltonian}. If we now solve the equation
$H(\eta_0,\eta_i) = 0$ for $\eta_0$, we obtain $\eta_0(\eta_i)$ and
plugging this in the action, while substituting integration over $t$
with $l_0$, we obtain
\begin{equation}
  \label{eq:quantization:action-reduced}
  S_\text{red}[(l_i,\eta_i)(l_0)] =
  \int \d l_0 \left[ \sum_i 2\,\eta_i\,\der{l_i}{l_0} + 2\,\eta_0(\eta_i)
  - \der{t}{l_0} \sum_{f \in F} \left( v_f\,C^f_\theta + w_f\,C^f_z \right)
  \right].
\end{equation}
We see that we have removed one pair of canonical coordinates, while
solving the Hamiltonian constraint. The Hamiltonian is now given by
$H_\text{red} = -2\,\eta_0$ together with the remaining constraints
and their Lagrange multipliers. Also, some of the complex constraints
will explicitly depend on the new time parameter $l_0$.

In theory, this method of introducing an internal time will thus yield
a system with non-trivial equations of motions. The explicit equations
of motion will depend on the solution of $\eta_0(\eta_i)$ from $H=0$,
which is in practice hard to solve.

The other way of implementing the constraints at the quantum level
comes down to first quantizing the complete phase space and then
constructing quantum mechanical operators $\hat{C}$ for each
constraint $C$ according to Dirac's prescription. A physical wave
function $\phi$ must then satisfy
\begin{equation}
  \label{eq:quantization:constraint-wavefunction}
  \hat{C} \, \phi = 0.
\end{equation}
Again, we are faced with the Hamiltonian constraint. If this
constraint is enforced at the quantum level, we have
$\hat{H} \, \phi = 0$, and thus that states are constant in time. If
we first solve the Hamiltonian constraint as in
\fref{eq:quantization:action-reduced}, we have a highly complicated
system, where a lot of symmetry has been removed.

\section{Quantization of the polygon model}
\label{sec:quantization:polygon}

Given the classical formulation of the polygon model in
\fref[plain]{chap:polygon}, we want to find a consistent choice of
quantization.

The most straightforward choice as proposed by 't Hooft in
\cite{'tHooft:1993nj}, is to promote the canonical coordinates
$l_i,\eta_j$ to operators, as in standard quantum mechanics. The
Poisson brackets of these variables are the same as in
\fref{eq:quantization:xp-commutator} (up to a factor $\mfrac{1}{2}$).

We now run into a problem: the Stone--Von Neumann uniqueness theorem
tells us that the canonical commutation relations
\fref[plain]{eq:quantization:xp-commutator} have one unique
representation, when $\hat{x}$ and $\hat{p}$ are self-adjoint
operators on a separable\footnote{%
  A topological space is separable when it has a dense subset that is
  countable. A Hilbert space of countable dimension is separable. If
  separability is not demanded, then other representations can be
  constructed, see \cite{Ashtekar:2002sn}. In these representations
  either $\hat{x}$ or $\hat{p}$ will not be a well-defined operator,
  although the exponentiated form will be well-defined. The spectrum
  of the other operator will still be the whole real line though.} %
Hilbert space (see e.g. \cite[p.~65]{Putnam:1967}). This
representation is the Heisenberg representation and implies that the
eigenvalue spectrum of both the $\hat{l}_i$ and $\hat{\eta}_j$ would
consist of the whole real line. The spectra of these operators in the
quantum theory do not correctly reflect some properties of the
original classical system, namely, the inequalities for the lengths
$l_i \ge 0$ and the triangle
inequalities~\fref[plain]{eq:polygon:triangle} for the boosts. We will
now try to analyze whether these apparent problems can be remedied.

\subsection{Transitions}

The inequalities that lengths should be positive can be directly
related to graph transitions. In the classical case, when a length
would become negative, we demand that a transition takes place
instead. This is because, at least classically, introducing negative
lengths might give rise to all sorts of strange things like negative
spatial volume or overlapping coordinate charts. These overlapping
coordinates would also show up when we ignore transition d in
\fref{fig:transitions}; a case that is not covered by the inequalities
$l_i \ge 0$.

The question is now how we want to treat this in a quantum theory.
Allowing negative lengths and ignoring transitions will clearly remove
the non-trivial dynamics from the model. Therefore it seems that we
should also restrict the quantum mechanical system to these length
inequalities and probably rule out overlapping coordinates like
in \fref{fig:polygon-intersect}, too.

If we try to implement transitions in the quantum theory, we face the
fact that transitions change the graph structure and the coordinates
of the phase space. In the case of a polygon merging or splitting
transition, they even change the dimension of phase space. This poses
a significant problem for the formulation of a quantized theory: how
should these different phase spaces be brought together in one quantum
theory?

In \cite{'tHooft:1993nj} 't Hooft suggests to make an analytical
continuation of the wave functions, such that they are continuous
across transitions. Implementing the constraints on the quantum level
should then also ensure that the wave functions are constant under
gauge transformations.

There are however not only the continuous symmetries generated by the
constraints, but there is also a discrete overcounting of physical
states: one and the same universe can be represented by multiple
completely different (on a graph structure level) tessellations. It
might be possible that these different graphs can be transformed into
one another by finite gauge transformations, but this is not a priori
clear.

This raises questions like whether one wants to keep those overcounted
graphs on the quantum level or not and how to deal with either option.
If one wants to remove this symmetry, then what should be the choice
of gauge fixing? If one wants to keep these discrete symmetries, how
do we construct one global Hilbert space and a proper measure on it?

For the first option, a possible way to partially solve the problem of
gauge fixing is offered in \cite{Kadar:2003ie}. There it is
conjectured that every element in the physical phase space can be
constructed as an OPT. This would allow one to write all multi-polygon
tessellations as OPT's by a suitable gauge transformation and thus
keep the phase space of fixed dimensions. Then one would either need
to explicitly find this `suitable gauge transformation', which seems a
non-trivial task, or one would have to show by other means that the
phase space is complete.

\subsection{The triangle inequalities}

In the construction of the polygon model, we imposed that there is no
cosmological constant or matter, which implies that space-time is
flat. These are sufficient conditions to obtain the triangle
inequalities.

As these inequalities are inherently connected to the way the model is
constructed, it seems that they should also be imposed on the quantum
level; otherwise, the quantum theory would have regions that are
classically forbidden and it is therefore difficult to see how a
correct classical limit could emerge. Furthermore the classical
Hamiltonian is only defined for those boosts that satisfy the triangle
inequalities.

Each triangle inequality in itself does not restrict a single boost
$\eta_i$ to a certain domain, but the inequalities do couple boosts at
a vertex: choosing values for two boosts at a vertex restricts the
value of the third boost to a bounded domain. Even stronger, these
inequalities couple between all boosts due to the fact that each edge
is connected to two vertices and all edges to each other through the
graph structure of the tessellation. This is in conflict with the
canonical commutation relations~\fref{eq:quantization:xp-commutator},
because by the Stone--Von Neumann theorem, each boost $\eta_i$ should
have spectrum $\R$ independently. This problem could be overcome, if
we can find a symplectic transformation\footnote{%
  A symplectic transformation is a transformation on a space, which
  has a symplectic form defined on it. That is in our case: a phase
  space with a Poisson bracket, where the symplectic transformation by
  definition preserves the symplectic form (thus the Poisson
  bracket).} %
of the phase space that decouples the inequalities for the new
coordinates.

\subsection{Other choices of quantum variables}

As laid out above, the coordinates $l_i,\eta_j$ are both restricted by
inequalities at the classical level, which gives problems when trying
to promote them to quantum operators. A possible way to avoid this
problem is to choose a different set of coordinates to be promoted to
quantum operators.

We can try to reparametrize our coordinates in such a way that the
inequality constraints simplify and/or decouple; or we might be able
to find a reparametrization, which exactly parametrizes the part of
phase space where the inequalities are satisfied.

These methods of choosing a different coordinatization of phase space
can be extended to a more general way to attack the problem of these
inequalities. We can choose a set of phase space functions and promote
these together with a quantum deformation of their Poisson algebra to
operators in a Lie algebra. This is not a fundamentally different
approach from the first; yet it does add the possibility of choosing
an enlarged set of phase space functions that are not independent.

\begin{figure}
  \center
  \includegraphics{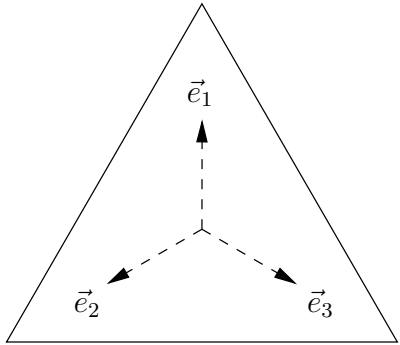}
  \caption{Coordinates used to parametrize a triangle.}
  \label{fig:triangle-params}
\end{figure}

We have tried to find a reparametrization of the boosts. The
restricted pairs of boost triples are shown in
\fref{fig:boosttriangle}. As a way to parametrize only the triples
satisfying the triangle inequality, we can first extract a common
scale factor $s$ from a boost triple. Then a triangle remains to be
parametrized. We can exploit its symmetry by choosing three coordinate
axes as in \fref{fig:triangle-params}. Now a point in this triangle
can be written as $\vec{x} = \sum_{i=1}^3 c_i\,\vec{e}_i$, but there
is a redundancy in this description, which can be lifted by imposing
the constraint $C = 1 - \sum_{i=1}^3 c_i \approx 0$. We also have to
restrict the range of the coefficients: $c_i \in \R_+$ suffices,
because it implies $c_i \in [0,1]$ together with the constraint.
Furthermore, a consistent quantization exists on $L^2(\R_+)$, see
\cite{Isham:1985bt}.

This approach does not solve all problems. The new variables are a
linear combination of the boosts, so the Poisson brackets will still
be of the same form. Since the new coordinates $c_i$ have a restricted
domain, the same problem will arise from the Stone--Von Neumann
theorem. Next to that, this way of parametrizing the boosts introduces
new variables $c_i$ per vertex, while each boost is connected to two
vertices. Thus, one would have to introduce constraints of the form
$c^1_i \approx c^2_j$ for each boost, where 1 and 2 label the two
vertices a boost is connected to.

Some remarks can be made about an approach in the way mentioned above.
We have seen that quantization using variables that obey the canonical
commutation relations gives problems with the eigenvalue spectrum.
There is also the problem that the constraints (and Hamiltonian) are
very complex non-polynomial expressions in terms of the boosts. This
will probably give rise to severe operator ordering ambiguities when
we try to implement these constraints at the quantum level, if
suitable quantum representations can be found at all.

To circumvent these problems, we can look for a set of phase space
functions that do not obey the canonical commutation relations and in
terms of which the constraints are functionally simpler. As an ansatz
we propose the functions $l_i, \sigma_j \!=\! \sinh(\eta_j)$ and
$\gamma_j \!=\! \cosh(\eta_j)$. These have the advantage that the
linearity of the constraints in the lengths is preserved and the
expressions obviously become simpler in terms of $\sigma_j, \gamma_j$.
The angles are given in terms of inverse (co)sines of rational
functions of the $\sigma_j, \gamma_j$ and thus not polynomial yet.
These functions obey the Poisson algebra
\begin{equation}
  \label{eq:quantization:poisson-l-sinh-cosh}
  \begin{split}
    \poisson{l_i}{\sigma_j}      &= \poissonleta \, \gamma_j,\\
    \poisson{l_i}{\gamma_j}      &= \poissonleta \, \sigma_j,\\
    \poisson{\sigma_i}{\gamma_j} &= 0,
  \end{split}
\end{equation}
which are not the canonical commutation relations anymore. We now want
to find a representation of this algebra such that $l$ and $\gamma$
have positive spectrum. We have not worked out whether choosing
$\hat{l}_i, \hat{\sigma}_j, \hat{\gamma}_j$ as quantum operators
yields a consistent quantization scheme though.

A different approach would be to only use variables associated with a
set of fundamental loops of the spatial manifold $\Sigma_g$. With each
such loop we can associate a Poincar\'e transformation, determined by
6 parameters and there are $2\,g\!-\!1$ independent loops. If we could
rewrite the polygon model in terms of these variables, there would
only be 3 constraints left. The $6(2\,g\!-\!1)$ variables are 6 more
than the $12(g\!-\!1)$ dimensions of the physical phase space; 3 of
these are constraints and the other 3 are gauge freedoms associated to
these constraints.

\section{Conclusions}
\label{sec:quantization:conclusions}

The original question whether space is quantized or time is
discretized seems to depend on the way one quantizes the theory. This
fact is well known, see e.g. \cite{Waelbroeck:1996sg} or
\cite[chap.~13]{Henneaux:1992ig}.

To draw conclusions about discreteness, one must first have a
consistent quantization scheme. In this chapter we saw, that the
obvious choice of canonical commutation relations has some serious
problems which cannot be overcome easily. Most notably the problem
that the spectrum of the basic quantum operators conflicts with the
classical length and boost constraints. Therefore it seems that one
should not draw conclusions about the spectra of length and time,
without first showing that a consistent quantization scheme can (at
least theoretically) be found.

%%% Local Variables: 
%%% mode: latex
%%% TeX-master: "thesis"
%%% End: 

%% file: constraintalgebra.tex
\chapter{The constraint algebra}
\label{chap:constraintalgebra}

\vspace{-3pt}
In this chapter, we will do some explicit calculations to determine
some properties of the algebra of Poisson brackets of the constraints.
Next, we try to interpret these constraints and the transformations
they generate.

The algebra of constraints expresses how a constraint transforms under
an infinitesimal gauge transformation with respect to another
constraint. In the case of first class constraints, the result should
be equal to zero on the constraint surface (a so-called `weak
equality'), because a classical solution must satisfy these
constraints and a gauge transformation preserves the solution and thus
the constraint too.

The calculation of the constraint algebra is interesting from
different perspectives. It allows us to check that these weak
equalities hold, which is a consistency check for the polygon model.
Furthermore, if one wants to quantize the model and implement the
constraints in the quantum theory, then this Poisson algebra should
have an analogue in terms of operator commutators in the quantum
theory. If they are not equal, this could give rise to quantum
anomalies \cite[p.~279]{Henneaux:1992ig}.

The constraints do not form a true Poisson algebra, because the
right-hand sides of the Poisson brackets turn out to be non-linear
expressions in terms of the original constraints and thus the linear
span of the constraints does not form a closed space under the Poisson
brackets. A true Poisson algebra is a vector space $X$ with
multiplication and a bi-linear and anti-symmetric Poisson bracket
\begin{equation}
  \label{eq:constraintalgebra:poissonalgebra}
  \poisson{.}{.} : X \times X \mapsto X,
\end{equation}
satisfying the Jacobi identity. Thus when we choose a basis $e_i$ of
$X$, we can write
\begin{equation}
  \label{eq:constraintalgebra:structconst}
  \poisson{e_i}{e_j} = f_{ij}^k \, e_k,
\end{equation}
where the $f_{ij}^k$ are the so-called structure constants.

Our algebra of constraints is not of this form, but can (in principle)
be rewritten in a form where the structure constants $f_{ij}^k$ are
replaced by structure functions $f_{ij}^k(l,\eta)$ on phase space. For
a proof, see \cite[p.~8~and~appendix~1.A]{Henneaux:1992ig}: the
Poisson brackets turn out to vanish on the constraint surface and can
thus be written as linear combinations of the constraints, with
coefficients given by structure functions. These structure functions
are not explicitly given however and might be very complicated.

\enlargethispage*{7pt}
\pagebreak

\section{Useful formulas}
\label{sec:constraintalgebra:formulas}

Before starting the calculation, we summarize the constraints and some
useful relations, which will be used to calculate the Poisson
brackets. For each polygon we have two constraints: the angular
constraint $C_\theta$ and the complex (closure) constraint $C_z$.
They are~\fref[plain]{eq:polygon:constraints}:
\begin{align*}
  C^A_\theta &= \sum_{j=1}^N (\pi - \alpha_j) - 2\,\pi,\\
  C^A_z      &= \sum_{j=1}^N l_j \exp( i \theta_j )
  \qquad\qquad\text{where}\qquad
  \theta_j    = \sum_{i=1}^j \pi - \alpha_i.
\end{align*}
The label $A = 1 \dots P$ labels the different polygons. The complex constraint
takes values in $\C$, so it is actually two real constraints. Instead
of working with the real and imaginary part, $C^A_z$ and $\bar{C}^A_z$
will be used to calculate all independent Poisson brackets.

Next we have the vertex relations~\fref{eq:polygon:vertexrelations},
from which the following identities can be derived:
\begin{align}
  \label{eq:constraintalgebra:dadn-eq}
  \pder{\alpha_a}{\eta_a} &= -2\,\frac{\sigma_a}{s_a\,\sigma_b\,\sigma_c},\\[5pt]
  \label{eq:constraintalgebra:dadn-ne}
  \pder{\alpha_a}{\eta_b} &= -2\,\frac{c_c\,\sigma_a}{s_a\,\sigma_b\,\sigma_c}
  = 2\,\frac{\sigma_b\,\gamma_c + c_a\,\sigma_c\,\gamma_b}{s_a\,\sigma_b\,\sigma_c},
\end{align}
where the indices $a,b,c$ label the three different edges and angles
at a vertex according to \fref{fig:vertex-params}. Note that these
results differ by a factor $2$ from those in \cite{Welling:1996hg}
(appearing from chain-rule differentiation of the $2\,\eta$
arguments).

The rightmost expression of~\fref{eq:constraintalgebra:dadn-ne}
contains only quantities that can be expressed in terms of boosts and
angles of a single polygon at that vertex. This can be used to derive
another useful relation between neighboring edges and angles along a
polygon:
\begin{align*}
&\nop e^{\pm i\alpha_{k+1}}\left(
   \pder{\alpha_{k+1}}{\eta_{k+1}} \mp 2\,i\,\frac{\gamma_{k+1}}{\sigma_{k+1}}\right)\\
&= 2\,(c_{k+1} \pm i\,s_{k+1})
    \frac{\sigma_{k+1}\,\gamma_k + c_{k+1}\,\sigma_k\,\gamma_{k+1}
                            \mp i\,s_{k+1}\,\sigma_k\,\gamma_{k+1}}
         {s_{k+1}\,\sigma_k\,\sigma_{k+1}}\\
&= 2\,\frac{c_{k+1}\,\gamma_k\,\sigma_{k+1} + \sigma_k\,\gamma_{k+1}}
            {s_{k+1}\,\sigma_k\,\sigma_{k+1}} \pm 2\,i\,\frac{\gamma_k}{\sigma_k}\\
&= \pder{\alpha_{k+1}}{\eta_k} \pm 2\,i\,\frac{\gamma_k}{\sigma_k}
\eqnumber\label{eq:constraintalgebra:angle-next}
\end{align*}

The relation $e^{i(\pi-\alpha_j)} = -e^{-i\alpha_j}$ is used
throughout the calculations; together with the formula above, we can
use it to simplify expressions a lot, as we will see.

As most of the work needed to calculate the different Poisson brackets
is very much alike, we will first derive two general expressions,
which will be used in further calculations.  Both expressions are
of the form
\begin{equation}
  \label{eq:constraintalgebra:dthdn}
  \sum_{k=1}^N e^{\pm i\Phi_k} \pder{\theta_j}{\nu_k},
\end{equation}
where we have two polygons and differentiate angles of the first
(indices $j$) with respect to boosts of the second (indices $k$).
$\theta_j, \Phi_k$ denote the angles of edges as
in~\fref{eq:polygon:theta}. This expression is calculated both for the
two polygons being equal and different. In these calculations we make
some assumptions~\fref[plain]{eq:constraintalgebra:assumptions}, which
can be found in the next section
\vpageref{eq:constraintalgebra:assumptions}.

First for $j,k$ indexing the same polygon, we have
\begin{align*}
&\nop \sum_{k=1}^N e^{\pm i\theta_k} \pder{\theta_j}{\eta_k}\\
&=-\sum_{k=1}^N e^{\pm i\theta_k} \sum_{l=1}^j \pder{\alpha_l}{\eta_k}\\
&=-\sum_{l=1}^j e^{\pm i\theta_{l-1}} \left(
  e^{\pm i\theta_N\,\delta_{l,1}}\pder{\alpha_l}{\eta_{l-1}}
 +e^{\pm i(\pi-\alpha_l)} \pder{\alpha_l}{\eta_l}\right)\\
\intertext{%
  using the fact that angles only depend on neighboring boosts. Notice
  the extra term $e^{\pm i\theta_N\,\delta_{l,1}}$: it is present
  because of the jump in complex angle factor, where the indices are
  taken cyclic modulo $N$ (but not sums, as noted in
  \fullref{eq:notation:cyclicindex}). This extra factor is 1 on the
  constraint surface, but non-trivial away from the constraint
  surface. Now we plug in the identity
  $0 = 2\,i\,(\mfrac{\gamma_{l}}{\sigma_{l}}-\mfrac{\gamma_{l}}{\sigma_{l}})$
  and shift the index of one of the terms to obtain}%
&=-\sum_{l=1}^j e^{\pm i\theta_{l-1}} \left(
  e^{\pm i\theta_N\,\delta_{l,1}}\left[
    \pder{\alpha_l}{\eta_{l-1}} \mp 2\,i\,\frac{\gamma_{l-1}}{\sigma_{l-1}}
  \right]
 +e^{\pm i(\pi-\alpha_l)}\left[
    \pder{\alpha_l}{\eta_l}     \pm 2\,i\,\frac{\gamma_l}{\sigma_l}
  \right]\right)\\
&\nop \pm 2\,i\left(
  e^{\pm i\theta_j}\,\frac{\gamma_j}{\sigma_j}
 -e^{\pm i\theta_N}\,\frac{\gamma_N}{\sigma_N}\right)\\
&= \pm 2\,i\left(
  e^{\pm i\theta_j}\,\frac{\gamma_j}{\sigma_j}
 -e^{\pm i\theta_N}\,\frac{\gamma_N}{\sigma_N}\right)
+ \left(1 - e^{\pm i\theta_N}\right)
  \left(\pder{\alpha_1}{\eta_N} \mp 2\,i\,\frac{\gamma_N}{\sigma_N}\right)\\
&= \pm 2\,i\left(
  e^{\pm i\theta_j}\,\frac{\gamma_j}{\sigma_j}-\frac{\gamma_N}{\sigma_N}\right)
+ \left(1 - e^{\pm i\theta_N}\right)\pder{\alpha_1}{\eta_N}.
\eqnumber\label{eq:constraintalgebra:dthdn-eq}
\end{align*}
In the second last step, we used
relation~\fref{eq:constraintalgebra:angle-next} to let all terms in
the summation cancel against each other, except for the boundary terms.

When the $j,k$ index different polygons, we get (again under the
assumptions~\fref[plain]{eq:constraintalgebra:assumptions}) only a
contribution from the common edge denoted by $j_c,k_c$. This reduces
the number of terms in the summation:
\begin{align*}
&\nop  \sum_{k=1}^N e^{\pm i\Phi_k} \pder{\theta_j}{\nu_k}\\
&= \sum_{k\in\{k_c-1,k_c,k_c+1\}}
     e^{\pm i\Phi_k} \times -\sum_{l=1}^j \pder{\alpha_l}{\nu_k}\\
&=-e^{\pm i\Phi_{k_c}} \sum_{k\in\{-1,0,1\}}
     \Big(-e^{\pm i(\beta_{k_c}  -\Phi_M\delta_{k_c,1})}\,\delta_{k,-1}
          +                                                 \delta_{k, 0}
          -e^{\mp i(\beta_{k_c+1}-\Phi_M\delta_{k_c,M})}\,\delta_{k, 1}\Big)\\
& \hspace{6cm} \times
     \left(\pder{\alpha_{j_c  }}{\nu_{k_c+k}}\delta_{j_c  \in[1,j]} +
           \pder{\alpha_{j_c+1}}{\nu_{k_c+k}}\delta_{j_c+1\in[1,j]}\right)\\
&=-e^{\pm i\Phi_{k_c}}\left[ \left( 
                                                      \pder{\alpha_{j_c  }}{\nu_{k_c  }}
        -e^{\mp i(\beta_{k_c+1}+\Phi_M\delta_{k_c,M})}\pder{\alpha_{j_c  }}{\nu_{k_c+1}}\right)
      \delta_{j_c  \in[1,j]}\right.\\
&\hspace{1.7cm}+\left.
  \left(                                              \pder{\alpha_{j_c+1}}{\nu_{k_c  }}
        -e^{\pm i(\beta_{k_c  }+\Phi_M\delta_{k_c,1})}\pder{\alpha_{j_c+1}}{\nu_{k_c-1}}\right)
      \delta_{j_c+1\in[1,j]}     \right]\\
&= 2\,e^{\pm i\Phi_{k_c}}\left[
  \left( \frac{c_{k_c+1}\,\sigma_{k_c+1}}{s_{j_c}\,\sigma_{j_c-1}\,\sigma_{j_c}}
        -e^{\mp i \Phi_M\delta_{k_c,M}}\,(c_{k_c+1} \mp i\,s_{k_c+1})
         \frac{           \sigma_{k_c+1}}{s_{j_c}\,\sigma_{j_c-1}\,\sigma_{j_c}}
      \right)\delta_{j_c  \in[1,j]}\right.\\
&\hspace{1.65cm}+\left.
  \left( \frac{c_{k_c}\,\sigma_{k_c-1}}{s_{j_c+1}\,\sigma_{j_c}\,\sigma_{j_c+1}}
        -e^{\pm i \Phi_M\delta_{k_c,1}}\,(c_{k_c  } \pm i\,s_{k_c  })
         \frac{         \sigma_{k_c-1}}{s_{j_c+1}\,\sigma_{j_c}\,\sigma_{j_c+1}}
      \right)\delta_{j_c+1\in[1,j]} \right]\\
\intertext{%
  where the term $e^{\pm i \Phi_M} = 1$ on the constraint
  surface and we split off $(1-e^{\pm i \Phi_M})\dots$, giving%
}
&= 2\,e^{\pm i\Phi_{k_c}}\left[
  \pm i\,\frac{s_{k_c+1}\,\sigma_{k_c+1}}
              {s_{j_c}  \,\sigma_{j_c-1}\,\sigma_{j_c}}\,\delta_{j_c  \in[1,j]}
  \mp i\,\frac{s_{k_c}  \,\sigma_{k_c-1}}
              {s_{j_c+1}\,\sigma_{j_c}\,\sigma_{j_c+1}}\,\delta_{j_c+1\in[1,j]}
\right.\\ &\hspace{2.1cm}\left.
  + \left(1-e^{\mp i \Phi_M\delta_{k_c,M}}\right)\,e^{\mp i\beta_{k_c+1}}\,
      \frac{\sigma_{k_c+1}}{s_{j_c}  \,\sigma_{j_c-1}\,\sigma_{j_c}}\,\delta_{j_c  \in[1,j]}
\right.\\ &\hspace{2.1cm}\left.
  + \left(1-e^{\pm i \Phi_M\delta_{k_c,1}}\right)\,e^{\pm i\beta_{k_c}}\,
      \frac{\sigma_{k_c-1}}{s_{j_c+1}\,\sigma_{j_c}\,\sigma_{j_c+1}}\,\delta_{j_c+1\in[1,j]}
\right]\\
&= \pm 2\,i\,e^{\pm i\Phi_{k_c}}\,\frac{1}{\sigma_{j_c}}
   \left(\delta_{j_c\in[1,j]} - \delta_{j_c+1\in[1,j]}\right)\\
&\nop - 2\,\left(1-e^{\pm i \Phi_M}\right)\,\frac{1}{s_{\beta_1}\,\sigma_{\nu_{k_c}}}\,
\Big[             \delta_{j_c+1\in[1,j]}\,\delta_{k_c,1} +
e^{\mp i\beta_1}\,\delta_{j_c  \in[1,j]}\,\delta_{k_c,M} \Big].
\eqnumber\label{eq:constraintalgebra:dthdn-ne}
\end{align*}
Most of the above calculation is a straightforward expansion of the
definitions. However one must keep good track of the indices and
cyclic boundary conditions, which make the result non-trivial: if
$j = N$ and $k \not\in \{1,M\}$ then the result reduces to zero
identically. Thus we see here already that the result depends on the
choice of labelling of the polygons from the $\delta_{k_c,1}$ and
$\delta_{k_c,M}$ terms.

\section{Calculation of the algebra}
\label{sec:constraintalgebra:calculation}

We want to calculate the complete Poisson algebra. All different,
independent Poisson brackets between the constraints are given by
\begin{equation}
  \label{eq:constraintalgebra:brackets}
  \poisson{C^A_\theta}{C^B_\theta},\;
  \poisson{C^A_\theta}{C^B_z},\;
  \poisson{C^A_z}{C^B_z},\;
  \poisson{C^A_z}{\bar{C}^B_z},
\end{equation}
where $A,B$ independently run over all polygons.  All other combinations
can be obtained by using anti-symmetry of the Poisson brackets or by
complex conjugation, for example:
\begin{equation*}
  \poisson{C^A_\theta}{\bar{C}^B_z} = \overline{\poisson{C^A_\theta}{C^B_z}}.
\end{equation*}

To start with, there are some constraints which are trivially equal to
zero. First of all, each constraint commutes with itself (when also
the polygon labels are identical: $A=B$). Secondly
\begin{equation}
  \label{eq:constraintalgebra:poisson-angle-angle}
  \poisson{C^A_\theta}{C^B_\theta} = 0
\end{equation}
because there is no dependence on the edge lengths in either
constraint.

Next we will calculate the Poisson brackets between all remaining
constraints under some non-generic assumptions. Similar calculations
were made previously in \cite{Welling:1996hg}. Here, we will perform
the calculation more explicitly and establish several corrections to
the results obtained in \cite{Welling:1996hg}. These results we will
then use as a starting point for calculating the general form of the
algebra, thus lifting the
assumptions~\fref[plain]{eq:constraintalgebra:assumptions}. 

The assumptions are as follows:
\begin{subequations}\label{eq:constraintalgebra:assumptions}
\begin{align}
& \text{\emph{no single polygon has two edges that have to be identified} and}\\
& \text{\emph{no two different polygons have more than one edge in common.}}
\end{align}
\end{subequations}
These assumptions are a theoretical, but also practical restriction,
as they do not cover the `one polygon tessellations' (OPT's, as in
\cite{Kadar:2003ie}). There every edge of the polygon has to be
identified with another edge of the same, single polygon. OPT's are
the simplest tessellations possible and therefore a practical set of
tessellations to parametrize phase space with. If we want to
explicitly use these OPT's as a starting point for a quantum theory,
we therefore need the full Poisson structure without the assumptions.

We will calculate the explicit expressions for the general Poisson
structure without these assumptions and show that the constraints do
also close in that case. This will be treated in
\fref[plain]{sec:constraintalgebra:generalizations} and a summary of
all results will be presented in \fref{sec:constraintalgebra:summary}.

When we calculate brackets between two different polygons (labeled I
and II), the result will be zero unless these polygons have an edge
in common. This is because the constraints only depend on variables
locally at the border of the polygon. Hence with our assumptions we
only have to look at cases where two different polygons have exactly
one edge in common.

\begin{figure}
  \center
  \includegraphics{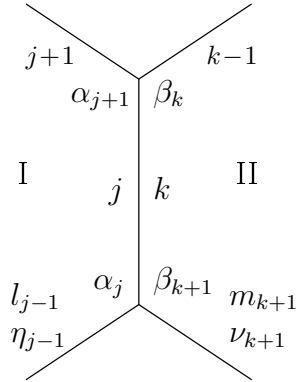}
  \caption{Label scheme for the calculation of the constraint algebra}
  \label{fig:constraint-params}
\end{figure}
In the calculations that follow, we will parametrize the two polygons
as in \fref{fig:constraint-params}. We label lengths, boosts and
angles of polygon I with $l,\eta,\alpha$ respectively and $j$ indexes
these. For polygon II we use $m,\nu,\beta$ with indices $k$. The
indices of the edges of polygon I and II that join each other will be
denoted by $j_c$ and $k_c$ respectively. Labelling will always be
counterclockwise oriented, see also \fref[plain]{chap:notation}.

We start by calculating the Poisson bracket of the angular constraint
with the complex constraint of the same polygon. As there is no
dependence on the lengths in the $C_\theta$, we only have to
differentiate it with respect to the boosts and thus $C_z$ only with
respect to the lengths. This yields
%%%%%%%%%%%%%%%%%%%%%%%%%%%%%%%%%%%%%%%%%
% Definition of poisson bracket result: %
\def\pangcomint#1{%
-\left(  \frac{\sigma_N\,\gamma_1 + c_1\,\sigma_1\,\gamma_N}{s_1\,\sigma_1\,\sigma_N}%
       -i\frac{\gamma_N}{\sigma_N}\right) \left(1 - e^{i\theta_N}\right)}
%%%%%%%%%%%%%%%%%%%%%%%%%%%%%%%%%%%%%%%%%
\begin{align*}
&\nop \poisson{C^{\rm I}_\theta}{C^{\rm I}_z}\\
&= \poisson{\theta_N}{\sum_{k=1}^N l_k\,e^{i\theta_k}}
\displaybreak\\
&=-\mfrac{1}{2} \sum_{k=1}^N e^{i\theta_k} \, \pder{\theta_N}{\eta_k}
\eqcomment{using definition \fref[plain]{eq:notation:poisson}}\\
&=-\mfrac{1}{2}\left(\pder{\alpha_1}{\eta_N} - 2\,i \frac{\gamma_N}{\sigma_N}\right)
  \left(1 - e^{i\theta_N}\right)
\eqcomment{using \fref[plain]{eq:constraintalgebra:dthdn-eq}}\\
&= \pangcomint{\rm I} \\
&= e^{i\theta_1} \left(  \frac{\sigma_1\,\gamma_N + c_1\,\sigma_N\,\gamma_1}{s_1\,\sigma_1\,\sigma_N}
                       +i\frac{\gamma_1}{\sigma_1}\right)
   \left(1 - e^{i\theta_N}\right) \approx 0.
\eqnumber\label{eq:constraintalgebra:poisson-angle-complex-eq}
\end{align*}
In the last step we used
equation~\fref{eq:constraintalgebra:angle-next} to rewrite the result
to the same form as in \cite{Welling:1996hg}. It coincides with the
latter expression up to a factor $2\,e^{i(\pi-\alpha_1)}$; the
exponential is due to a different definition of the complex constraint
and the factor $2$ probably due to a calculation error, as noted below
\fref{eq:constraintalgebra:dadn-ne}.

When we take the complex constraint together with its complex
conjugate, we get terms from both differentiating the first with
respect to the boosts and the second with respect to the lengths and
vice versa:
%%%%%%%%%%%%%%%%%%%%%%%%%%%%%%%%%%%%%%%%%
% Definition of poisson bracket result: %
\def\pcomcbrint#1{%
 \frac{\gamma_N}{\sigma_N} \left(C^{#1}_z - \bar{C}^{#1}_z\right)%
-\frac{i}{2}\pder{\alpha_1}{\eta_N}%
  \Big[   C ^{#1}_z\,\left(1-e^{-i\theta_N}\right)%
    +\bar{C}^{#1}_z\,\left(1-e^{ i\theta_N}\right) \Big]}
%%%%%%%%%%%%%%%%%%%%%%%%%%%%%%%%%%%%%%%%%
\begin{align*}
&\nop \poisson{C^{\rm I}_z}{\bar{C}^{\rm I}_z} \\
&= \poisson{\sum_{j=1}^N l_j\,e^{i\theta_j}}{\sum_{k=1}^N l_k\,e^{-i\theta_k}}\\
&= \mfrac{1}{2} \sum_{j,k=1}^N \left(
   l_k\,e^{i\theta_j}\,e^{-i\theta_k}\times -i\,\pder{\theta_k}{\eta_j} \; - \;
   l_j\,e^{i\theta_j}\,e^{-i\theta_k}\times  i\,\pder{\theta_j}{\eta_k} \right)\\
&=-\frac{i}{2}\sum_{k=1}^N l_k\,e^{-i\theta_k}\left[
  + 2\,i\left(e^{ i\theta_k}\frac{\gamma_k}{\sigma_k}-\frac{\gamma_N}{\sigma_N}\right)
  + \left(1-e^{ i\theta_N}\right)\pder{\alpha_1}{\eta_N}\right]\\
&\nop
  -\frac{i}{2}\sum_{j=1}^N l_j\,e^{ i\theta_j}\left[
  - 2\,i\left(e^{-i\theta_j}\frac{\gamma_j}{\sigma_j}-\frac{\gamma_N}{\sigma_N}\right)
  + \left(1-e^{-i\theta_N}\right)\pder{\alpha_1}{\eta_N}\right]
\eqcomment{using \fref[plain]{eq:constraintalgebra:dthdn-eq}}\\
&= \pcomcbrint{\rm I} \\
&= 2\,i\,\frac{\gamma_N}{\sigma_N}\Im\Big(C^{\rm I}_z\Big)%
    - i\,\pder{\alpha_1}{\eta_N}\,\Re\Big(C^{\rm I}_z\left(1-e^{-i\theta_N}\right)\Big)
\approx 0.
\eqnumber\label{eq:constraintalgebra:poisson-complex-complexbar-eq}
\end{align*}
This result is somewhat different than that in \cite{Welling:1996hg}:
in the first term we have a relative minus sign instead of plus and
the second ``boundary'' term is added; it arises due to the cyclicity
of the edge and angle indices, but one has to be careful with the
interpretation of sums of angles as noted in
\fref{eq:notation:cyclicindex}.

Notice that the correctness of the minus sign is easily checked by
complex conjugating the complete Poisson brackets, because is should
satisfy
\begin{equation}
  \overline{\poisson{C_z}{\bar{C}_z}}
= \poisson{\bar{C}_z}{C_z}
=-\poisson{C_z}{\bar{C}_z}.
\end{equation}
We can check that the second term does also obey this condition.

Next we have the Poisson brackets between two different
polygons. Using equation~\fref{eq:constraintalgebra:dthdn-ne} we get
%%%%%%%%%%%%%%%%%%%%%%%%%%%%%%%%%%%%%%%%%
% Definition of poisson bracket result: %
\def\pangcomext#1#2{%
\left(1-e^{i \Phi_M}\right)\,\frac{1}{s_{\beta_1}\,\sigma_{\nu_{k_c}}}\,%
\left(\delta_{k_c,1} + e^{-i\beta_1}\,\delta_{k_c,M}\right)}
%%%%%%%%%%%%%%%%%%%%%%%%%%%%%%%%%%%%%%%%%
\begin{align*}
&\nop \poisson{C^{\rm I}_\theta}{C^{\rm II}_z} \\
&= \poisson{\theta_N}{\sum_{k=1}^M m_k \, e^{i\Phi_k}} \\
&=-\mfrac{1}{2} \sum_{k=1}^M e^{i\Phi_k} \pder{\theta_N}{\nu_k}\\
&= -i\,e^{i\Phi_{k_c}}\,\frac{1}{\sigma_{j_c}}
     \left(\delta_{j_c\in[1,N]} - \delta_{j_c+1\in[1,N]}\right)\\
&\nop +\left(1-e^{i \Phi_M}\right)\,\frac{1}{s_{\beta_1}\,\sigma_{\nu_{k_c}}}\,
\left(               \delta_{j_c+1\in[1,j]}\,\delta_{k_c,1}
    + e^{-i\beta_1}\,\delta_{j_c  \in[1,j]}\,\delta_{k_c,M} \right)\\
&= \pangcomext{\rm I}{\rm II} \approx 0
\eqnumber\label{eq:constraintalgebra:poisson-angle-complex-ne}
\end{align*}
by straightforward calculation, where we have to remember that the
indices are cyclic and thus trivially
$\delta_{j_c\in[1,N]} - \delta_{j_c+1\in[1,N]} = 1 - 1 = 0$. Comparing
again with \cite{Welling:1996hg}, we have found an additional boundary
term arising from cyclicity of the indices along the polygon.

\pagebreak

The calculation of Poisson brackets between two complex constraints gives
%%%%%%%%%%%%%%%%%%%%%%%%%%%%%%%%%%%%%%%%%
% Definition of poisson bracket result: %
\def\pcomcomext#1#2#3{%
-\,\frac{e^{i\Phi_{k_c}  }}{\sigma_{k_c}}\,C^{#1}_z\,\delta_{j_c,N}%
+  \frac{e^{i\theta_{j_c}}}{\sigma_{j_c}}\,C^{#2}_z\,\delta_{k_c,M}\\%
&\nop \hspace{#3}%
-\,\frac{i}{s_{\alpha_1}\,\sigma_{\eta_{j_c}}}\left(1-e^{i\theta_N}\right)%
  \Big[                Z^{#2}_{k_c+1}\,\delta_{j_c,1}%
      +e^{-i\alpha_1}\,Z^{#2}_{k_c  }\,\delta_{j_c,N}\Big]\\%
&\nop \hspace{#3}%
+\,\frac{i}{s_{\beta_1}\,\sigma_{\nu_{k_c}}}\left(1-e^{i\Phi_M}\right)%
  \Big[                Z^{#1}_{j_c+1}\,\delta_{k_c,1}%
      +e^{-i\beta_1}\, Z^{#1}_{j_c  }\,\delta_{k_c,M}\Big]}
%%%%%%%%%%%%%%%%%%%%%%%%%%%%%%%%%%%%%%%%%
\begin{align*}
&\nop \poisson{C^{\rm I}_z}{C^{\rm II}_z} \\
&= \poisson{\sum_{j=1}^N l_j\,e^{i\theta_j}}{\sum_{k=1}^M m_k\,e^{i\Phi_k}}\\
&= \frac{i}{2}\sum_{j,k} \left(
  m_k\,e^{i\theta_j}\,e^{i\Phi_k} \pder{\Phi_k}{\eta_j}
 -l_j\,e^{i\theta_j}\,e^{i\Phi_k} \pder{\theta_j}{\nu_k} \right)\\
&= \sum_{k=1}^M m_k\,e^{i\Phi_k}\left[\rule{0cm}{0.8cm}
  -\frac{e^{i\theta_{j_c}}}{\sigma_{k_c}}
     \left(\delta_{k_c\in[1,k]} - \delta_{k_c+1\in[1,k]}\right)\right.\\
&\nop \hspace{2.25cm} \left.\rule{0cm}{0.8cm}
  -\frac{i}{s_{\alpha_1}\,\sigma_{\eta_{j_c}}}\left(1-e^{i\theta_N}\right)
    \left(              \delta_{k_c+1\in[1,k]}\,\delta_{j_c,1} +
          e^{-i\alpha_1}\delta_{k_c  \in[1,k]}\,\delta_{j_c,N}\right)
  \right]\\
&\nop - \; \{l,\eta,\theta,j\} \leftrightarrow \{m,\nu,\Phi,k\}
\eqcomment{inserting \fref[plain]{eq:constraintalgebra:dthdn-ne}}\\
&= \pcomcomext{\rm I}{\rm II}{0pt} \approx 0,
\eqnumber\label{eq:constraintalgebra:poisson-complex-complex-ne}
\end{align*}
where we have defined
\begin{equation}
  \label{eq:constraintalgebra:partial-complex}
  Z^A_{k} = \sum_{j=k}^{N} l_j\,e^{i\theta_j}
\end{equation}
as the partial complex constraint of polygon $A$ from edge
$k$. Also, we used the relations
\begin{equation}
  \label{eq:constraintalgebra:delta-relations}
  \delta_{j_c\in[1,j]} - \delta_{j_c+1\in[1,j]} = \delta_{j,j_c} - \delta_{j_c,N}
  \qquad\text{and}\qquad
  \delta_{j_c\in[1,j]} = \delta_{j\in[j_c,N]}.
\end{equation}

Completely analogously, we find for a complex and complex-conjugate
constraint
%%%%%%%%%%%%%%%%%%%%%%%%%%%%%%%%%%%%%%%%%
% Definition of poisson bracket result: %
\def\pcomcbrext#1#2#3{%
+\,\frac{e^{-i\Phi_{k_c} }}{\sigma_{k_c}}\,      C^{#1}_z\,\delta_{j_c,N}%
-  \frac{e^{i\theta_{j_c}}}{\sigma_{j_c}}\,\bar{C}^{#2}_z\,\delta_{k_c,M}\\%
&\nop \hspace{#3}%
+\,\frac{i}{s_{\alpha_1}\,\sigma_{\eta_{j_c}}}\left(1-e^{i\theta_N}\right)%
  \Big[                \bar{Z}^{#2}_{k_c+1}\,\delta_{j_c,1}%
      +e^{-i\alpha_1}\,\bar{Z}^{#2}_{k_c  }\,\delta_{j_c,N} \Big]\\%
&\nop \hspace{#3}%
+\,\frac{i}{s_{\beta_1}\,\sigma_{\nu_{k_c}}}\left(1-e^{-i\Phi_M}\right)%
  \Big[                      Z^{#1}_{j_c+1}\,\delta_{k_c,1}%
      +e^{i\beta_1} \,       Z^{#1}_{j_c  }\,\delta_{k_c,M} \Big]}
%%%%%%%%%%%%%%%%%%%%%%%%%%%%%%%%%%%%%%%%%
\begin{align*}
&\nop \poisson{C^{\rm I}_z}{\bar{C}^{\rm II}_z} \\
&= \pcomcbrext{\rm I}{\rm II}{0pt} \approx 0.
\eqnumber\label{eq:constraintalgebra:poisson-complex-complexbar-ne}
\end{align*}
Both of these Poisson brackets yield expressions that vanish on the
constraint surface, but do not identically vanish on the whole phase
space, unlike the results in \cite{Welling:1996hg}: when one of the
indices $j_c, k_c$ is equal to the first or last of its polygon, the
result is non-zero.

As already pointed out in \cite{'tHooft:1993nj}, the Hamiltonian flow
generated by the complex constraints should (on shell) be related to
Lorentz transformations of the coordinate frames of the polygons. This
is a residual gauge freedom that is left after fixing the constraints
$\mathcal{H}^i$ in \fref{eq:representation:mom-constr}. With this
interpretation, one would not expect the Poisson brackets between
these constraints to vanish identically (as is the result in
\cite{Welling:1996hg}): the complex constraint should be related to
the constraints $\mathcal{H}^i$ and their corresponding gauge freedom
to the Lorentz transformations of the local frame of the polygon.

\begin{figure}
  \center
  \includegraphics{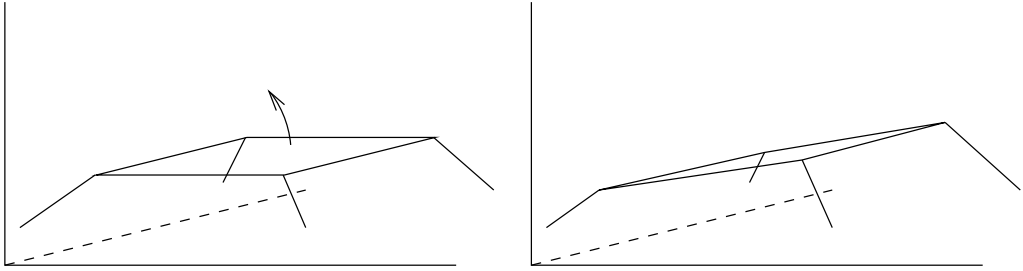}
  \caption{Rotation of the plane of a polygon and its action on edge
    lengths and extrinsic curvature surrounding the polygon.}
  \label{fig:polygon-rotation}
\end{figure}
\begin{figure}
  \centering
  \begin{minipage}[t]{.50\textwidth}
    \centering
    \includegraphics{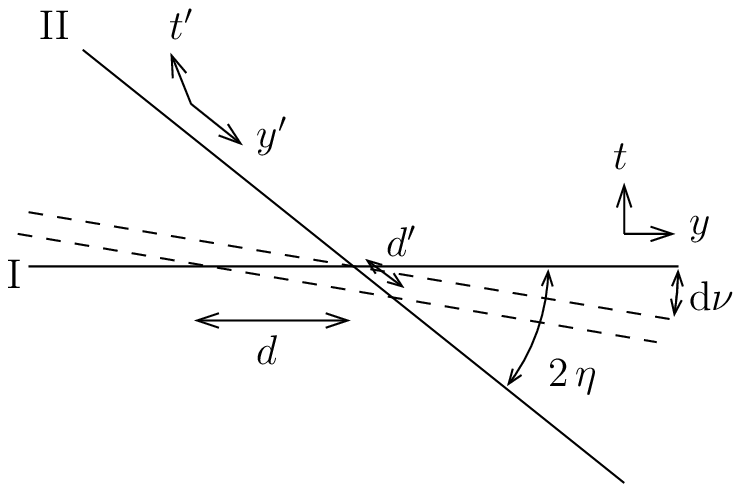}
    \caption{Action of a Lorentz transformation on a complex
      constraint. $(y,t)$ and $(y',t')$ denote the coordinates within
      polygons I and II respectively.}
    \label{fig:lorentz-constraint}
  \end{minipage}
  \hfill
  \begin{minipage}[t]{.44\textwidth}
    \centering
    \includegraphics{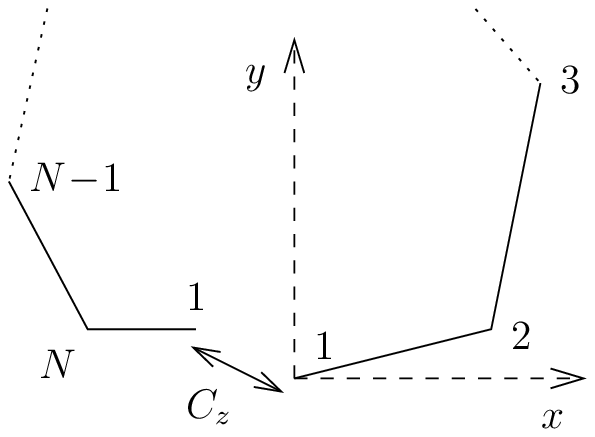}
    \caption{Polygon with non-closing complex constraint: vertex 1
      shows up on both sides of the deficit $C_z \ne 0$.}
    \label{fig:constraint-deficit}
  \end{minipage}
\end{figure}

Suppose now that the complex constraint in polygon I does not close
by an amount $d$ in the $y$-direction ($C_z^{\rm I} = i\,d \neq 0$)
and we generate a Lorentz transformation in polygon I. This Lorentz
transformation can be seen as a ``space-time rotation'' of the plane
of polygon I as in \fref{fig:polygon-rotation}: the (infinitesimal)
boost of the Lorentz transformation corresponds to an angle that the
plane of polygon I is rotated by. This will affect the lengths and
boosts of all edges along that polygon and also the lengths of edges
connected to those. We now consider the effect of this transformation
on a neighboring polygon II. We suppose that they have an edge in
common, which has indices $j_c, k_c$ in polygons I and II
respectively, where we choose $j_c = N, k_c = M$. The implicit
assumption is made that the closure deficit $C_z^{\rm I}$ is located
between edges $j = N$ and $j = 1$. Now the position of vertex 1 (which
is located between edges $j = N$ and $j = 1$) depends on whether we
view it as the first or last point of the constraint $C^{\rm I}_z$,
see \fref{fig:constraint-deficit}, where we have implicit complex
coordinates $z = x + i\,y$ and the origin is placed at vertex 1 viewed
as starting point of the complex constraint.

This will give rise to a different change of the intersection line
under this Lorentz transformation and thus a change in $C^{\rm II}_z$.
\Fref{fig:lorentz-constraint} shows the geometrical picture in the
$(y,t)$-plane. The lines are the planes of polygons I and II and their
intersection point is the common edge. Now vertex 1 shows up as this
intersection point, but also at a distance $d$ apart.

The action of an infinitesimal Lorentz transformation of polygon I by
an amount $\d\nu$ is shown as the dashed lines. The upper one is
rotated around the origin of polygon I (which is vertex 1) and the
lower one around vertex 1 as viewed after going around polygon I,
after which it is shifted by a distance $d$. These generate different
new intersection points between polygons I and II. We can now
calculate the distance between those intersection points explicitly,
which yields
\begin{equation}
  \label{eq:constraintalgebra:1-gen-lorentz}
  d' = \frac{d}{\sinh(2\,\eta)}\,\d \nu.
\end{equation}
This matches the result
\begin{equation}
  \label{eq:constraintalgebra:2-gen-lorentz}
  \poisson{C^{\rm II}_z}{C^{\rm I}_z}
  = \frac{e^{i\Phi_{k_c}}}{\sigma_{k_c}}\,C^{\rm I}_z\,\delta_{j_c,N}
  = \frac{i\,d}{\sigma},
\end{equation}
where we have inserted that all constraints close, except
$C^{\rm I}_z = i\,d$ and thus all other terms from
\fref{eq:constraintalgebra:poisson-complex-complex-ne} vanish and
$\Phi_M = 2\,\pi$.

Thus this geometrical viewpoint seems to indicate that indeed the
Poisson bracket $\poisson{C^{\rm I}_z}{C^{\rm II}_z}$ is not
identically zero outside the constraint surface, if we interpret the
complex constraints as generators of Lorentz transformations.

\pagebreak

\section{Generalizations}
\label{sec:constraintalgebra:generalizations}

We can generalize the
assumptions~\fref[plain]{eq:constraintalgebra:assumptions} made in the
previous section to the arbitrary case, which will recover the
explicit full Poisson structure. We will take two different
approaches: the first one is to consider the Poisson brackets of the
constraints with the Hamiltonian, using the Hamiltonian flow
in~\fref{eq:polygon:dlength}. The Hamiltonian is the sum of the
angular constraints, so for an OPT we have $H = C_\theta$. This will
then give us the Poisson bracket $\poisson{C_\theta}{C_z}$ in the case
of an OPT without any assumptions made. The second approach will be to
lift the assumptions in a general setting. The results of these
approaches are then shown to match in the case of an OPT.

\subsection{The case of an OPT}

To calculate the Poisson bracket $\poisson{H}{C_z}$, we first need a
formula relating the growth of edges along a polygon. This formula and
its derivation is analogous to~\fref{eq:constraintalgebra:dthdn-ne}.
However we note that in the definition of the $\sigma, \gamma$
in~\fref{eq:constraintalgebra:angle-next-full}, the factor $2$ is
missing with respect to our original
definition~\fref[plain]{eq:notation:boost-shorthand}, so
$\sigma_i = \sinh(\eta_i), \gamma_i = \cosh(\eta_i)$. We have
\begin{align*}
  \label{eq:constraintalgebra:angle-next-full}
&\nop e^{\pm i\alpha_{k+1}}\left(
   \dot{l}_{k+1,1} \pm i\,\frac{\sigma_{k+1}}{\gamma_{k+1}}\right)\\
&=-(c_{k+1} \pm i\,s_{k+1})
   \frac{\sigma_k\,\gamma_{k+1} + c_{k+1}\,\sigma_{k+1}\,\gamma_k
                           \mp i\,s_{k+1}\,\sigma_{k+1}\,\gamma_k}
        {s_{k+1}\,\gamma_k\,\gamma_{k+1}}\\
&=-\frac{\sigma_{k+1}\,\gamma_k + c_{k+1}\,\sigma_k\,\gamma_{k+1}}
        {s_{k+1}\,\gamma_k\,\gamma_{k+1}} \mp i\,\frac{\sigma_k}{\gamma_k}\\
&= \dot{l}_{k,2} \mp i\,\frac{\sigma_k}{\gamma_k}. \eqnumber
\end{align*}

We can use the previous relation to rewrite $\poisson{C_z}{H}$ (the
time evolution of the complex constraint) in the same way as we did
with~\fref{eq:constraintalgebra:poisson-angle-complex-eq}:
%%%%%%%%%%%%%%%%%%%%%%%%%%%%%%%%%%%%%%%%%
% Definition of poisson bracket result: %
\def\phamcom{%
  \left(-\frac{v_N\,c_1 + v_1}{s_1} + i\,v_N\right)%
  \left(e^{i\theta_N} - 1\right)}
%%%%%%%%%%%%%%%%%%%%%%%%%%%%%%%%%%%%%%%%%
\begin{align*}
&\nop \poisson{C_z}{H}\\
&= \poisson{\sum_{j=1}^N l_j\,e^{i\theta_j}}{H}\\
&= \sum_{j=1}^N e^{i\theta_j}\,\dot{l}_j\\
&= \sum_{j=1}^N e^{i\theta_j}\left(
  \dot{l}_{j,1} - i\,\frac{\sigma_j}{\gamma_j}
 +\dot{l}_{j,2} + i\,\frac{\sigma_j}{\gamma_j}
  \right)
\displaybreak\\
&= e^{i\theta_1} \left(\dot{l}_{1,1} - i\,\frac{\sigma_1}{\gamma_1}\right)
   \left(1 - e^{i\theta_N}\right)
   \eqcomment{using \fref[plain]{eq:constraintalgebra:angle-next-full}}\\
&= \left(\dot{l}_{N,2} + i\,\frac{\sigma_N}{\gamma_N}\right)
   \left(e^{i\theta_N} - 1\right)\\
&= \phamcom \approx 0
\eqnumber\label{eq:constraintalgebra:poisson-hamil-complex}
\end{align*}
Here we recognize the left part of the final expression as minus the
velocity of vertex 1 in the coordinate-frame of the polygon (in
complex notation, as was used for the formulation of the complex
constraint). The factor $e^{i\theta_N} - 1$ precisely gives the
difference of viewing this point as the first or the last when going
around the polygon. Thus this intuitively matches the expected result
of the time evolution of the closure constraint.

\subsection{The case of a general tessellation}

We now want to generalize the calculated Poisson brackets between two
polygons to the arbitrary case without
assumptions~\fref[plain]{eq:constraintalgebra:assumptions}.

The underlying idea is as follows: we can split the Poisson bracket
expression into parts (see \fref{eq:constraintalgebra:poisson-split}
for a precise specification), taking account for each common edge
separately.  For this, we first have to show that there are some
restrictions to the ways two polygons can have multiple edges in
common, such that we can make such a splitting. For this to work,
different common edges should not interfere with each other in the
calculation of the Poisson brackets.

\begin{figure}[b]
  \center
  \includegraphics{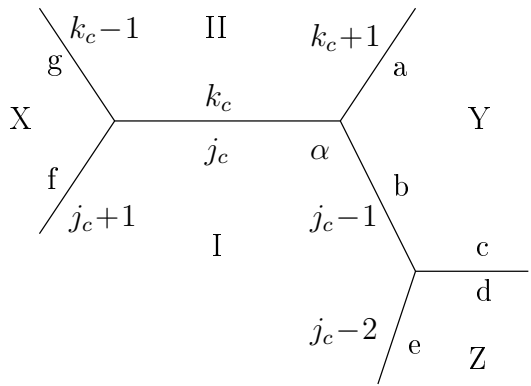}
  \caption{Multiple common edges between two polygons.}
  \label{fig:constraint-common-edges}
\end{figure}

To show that the ``condition'' above is fullfilled, let us consider
two polygons as in \fref{fig:constraint-common-edges}, which have at
least an edge in common for edge indices $j_c,k_c$. The polygons X, Y
and Z can be copies of polygons I or II and as such can give rise to
multiple common edges.

The only non-vanishing terms in the Poisson brackets appear from the
differentiation of one term in a complex (conjugate) constraint (say
polygon II) to the edge length and then the other constraint (polygon
I) differentiated with respect to the corresponding boost. This means
differentiation of an angle of polygon I, as the angles are the only
quantities in the constraints that depend on boosts.

If two common edges cannot appear in such a way that their `split'
Poisson brackets would both contain a specific angle term of polygon I
being differentiated to a boost of the same edge of polygon II, we can
replace the Poisson brackets by two separate Poisson brackets where we
treat the common edges one at a time. Note that here an `edge' is a
boundary element of a polygon, so each pair $(l_i, \eta_i)$ is
assigned to two edges, when the edges are viewed as boundaries of
polygons.

Thus the ``condition'' above can be stated in a more precise way as
follows:
\begin{equation}\label{eq:constraintalgebra:generalize-condition}
  \parbox{0.85\textwidth}{\emph{each edge from polygon II has at most
      one way in which it is connected to each angle of polygon I (and
      the converse, which follows from interchanging polygons I and II).}}
\end{equation}

The proof of this statement hinges on the fact that the graph of the
tessellation has trivalent vertices. We will first give an insightful
proof followed by a more detailed proof.

The statement can be seen to hold in the following way. If an angle
would be connected twice to an edge, then this edge has its endpoints
on one vertex. Because of trivalency of the vertices, both ends of
that edge must be neighboring at that vertex and thus the edge
encloses a polygon in itself. This is clearly a contradiction.

For the detailed proof, we start off with the case of an angle of
polygon I and an edge of polygon II. Proving
\fref{eq:constraintalgebra:generalize-condition} boils down to a
tedious but straightforward treatment of all possible scenario's. With
some symmetry arguments, we can however reduce this a bit. Suppose
that an angle of polygon I is connected twice to an edge of polygon
II. It must be an angle at one side of a common edge. Because of
symmetry we can take angle $\alpha$ as shown in
\fref{fig:constraint-common-edges} without loss of generality.

Now edge $k_c$ cannot be connected to $\alpha$ more than once, because
it is already joined with edge $j_c$ and thus cannot be joined with
$j_c-1$ and it cannot be equal to edge $k_c+1$. Also edge $k_c+1$
cannot be connected more than once, because if it would be connected
to $j_c-1$ (as $b$), then $k_c$ would be connected to $k_c+1$ (as
$a$). Finally, edges $a$ and $b$ obviously cannot be equal, so
condition~\fref[plain]{eq:constraintalgebra:generalize-condition} is
fullfilled for angles of polygon I.

To apply the idea as outlined above to the generalization of the
Poisson brackets, we first have to explicitly define what we mean by
a splitting of a Poisson bracket into separate parts for each common
edge. Let us denote the pairs of common edges of polygon $A$ and $B$
by $(j_c,k_c)_{c=1..n_c}$. We now introduce the notation
\begin{equation}
  \label{eq:constraintalgebra:poisson-split}
  \poisson{C^A}{C^B}_0
  \qquad\text{and}\qquad
  \poisson{C^A}{C^B}_c.
\end{equation}
The left expression shall denote the Poisson brackets of two
constraints, where no edges are identified. This expression will thus
only be non-zero when $A = B$ and then yields the `internal terms'.
The right expression shall denote the Poisson brackets of these
constraints, where no internal contributions are counted (so we treat
$A \ne B$) and only the edge pair $(j_c,k_c)$ is treated as common;
the others are treated as not being common. Thus by their definitions,
these expressions are equal to the corresponding expression calculated
in \fref[plain]{sec:constraintalgebra:calculation} under the
assumptions~\fref{eq:constraintalgebra:assumptions}:
$\poisson{C^A}{C^B}_0$ for the case $A = B$ and $\poisson{C^A}{C^B}_c$
for the case $A \ne B$. We will generalize the Poisson brackets
already calculated to arbitrary conditions, by proving that one can
split the general Poisson brackets as
\begin{equation}
  \label{eq:constraintalgebra:poisson-gen-split}
  \poisson{C^A}{C^B} =
  \poisson{C^A}{C^B}_0 + \sum_{c=1}^{n_c} \poisson{C^A}{C^B}_c.
\end{equation}

Given two polygons I and II, we denote their pairs of common edges by
$(j_c,k_c)_{c=1..n_c}$ and first investigate the Poisson brackets
$\poisson{\theta^{\rm I}_j}{C^{\rm II}_z}$:
\begin{align*}
&\nop \sum_{c=1}^{n_c} \poisson{\theta^{\rm I}_j}{C^{\rm II}_z}_c\\
&= \sum_{c=1}^{n_c} \Big[
  \poisson{-\alpha_{j_c  }\,\delta_{j_c  \in[1,j]}}
          {m_{k_c-1}\,e^{i\Phi_{k_c-1}}+m_{k_c  }\,e^{i\Phi_{k_c  }}} +\\
&\nop \hspace{0.95cm}
  \poisson{-\alpha_{j_c+1}\,\delta_{j_c+1\in[1,j]}}
          {m_{k_c  }\,e^{i\Phi_{k_c  }}+m_{k_c+1}\,e^{i\Phi_{k_c+1}}}\Big]\\
&= \sum_{j'=1}^j \poisson{-\alpha_{j'}}{C^{\rm II}_z}\\
&= \poisson{\theta^{\rm I}_j}{C^{\rm II}_z}
\eqnumber\label{eq:constraintalgebra:poisson-theta-complex-gen-ne}
\end{align*}
Working backward from the desired result, we first only write down
those terms that will contribute for each common edge. In the next
step, we use the fact that each combination of angle $\alpha$ of
polygon I and edge $m$ of polygon II does show up not more than once
in the full Poisson bracket, as stated in
\fref{eq:constraintalgebra:generalize-condition}. It also does show up
exactly once, as it is part of a single common edge bracket. All
remaining terms in $\poisson{\theta^{\rm I}_j}{C^{\rm II}_z}$ are zero
now, because we have exactly taken into account all terms contributing
to all common edge Poisson brackets and there are no internal
contributions, because we considered two different polygons.

\clearpage
In the case that both constraints are from the same polygon, we can 
essentially repeat the argument, except that we have to add the terms
that arise from the internal contributions:
\begin{align*}
&\nop            \poisson{\theta^{\rm I}_j}{C^{\rm I}_z}_0 +
\sum_{c=1}^{n_c} \poisson{\theta^{\rm I}_j}{C^{\rm I}_z}_c\\
&= \sum_{k=1}^N \poisson{-\alpha_k\,\delta_{k\in[1,j]}-\alpha_{k+1}\,\delta_{k+1\in[1,j]}}
                         {l_k\,e^{i\theta_k}} +\\
&\nop \sum_{c=1}^{n_c} \Big[
  \poisson{-\alpha_{j_c  }\,\delta_{j_c  \in[1,j]}}
          {l_{k_c-1}\,e^{i\theta_{k_c-1}}+l_{k_c  }\,e^{i\theta_{k_c  }}} +\\
&\nop \hspace{0.95cm}
  \poisson{-\alpha_{j_c+1}\,\delta_{j_c+1\in[1,j]}}
          {l_{k_c  }\,e^{i\theta_{k_c  }}+l_{k_c+1}\,e^{i\theta_{k_c+1}}}\Big]\\
&= \sum_{j'=1}^j \poisson{-\alpha_{j'}}{C^{\rm II}_z}\\
&= \poisson{\theta^{\rm I}_j}{C^{\rm II}_z}.
\eqnumber\label{eq:constraintalgebra:poisson-theta-complex-gen-eq}
\end{align*}
The extra terms do not change the argument, because also with these
added, condition~\fref{eq:constraintalgebra:generalize-condition}
still holds and all terms that we started with are exactly those that
will be present in the full Poisson bracket.

Now that we have shown in
\fref[plain]{eq:constraintalgebra:poisson-theta-complex-gen-ne} and
\fref[plain]{eq:constraintalgebra:poisson-theta-complex-gen-eq} that a
split can be made for the brackets $\poisson{\theta^A_j}{C^B_z}$ for
all $A,B$, we can easily extend the result to all possible Poisson
brackets. The application of
\fref[plain]{eq:constraintalgebra:poisson-theta-complex-gen-ne} to
$\poisson{C_\theta^{\rm I}}{C_z^{\rm II}}$ is trivial by plugging in
$j=N$ and yields
\begin{align*}
&\nop \poisson{C_\theta^{\rm I}}{C_z^{\rm II}}\\
&= \sum_{c=1}^{n_c} \poisson{C_\theta^{\rm I}}{C_z^{\rm II}}_c
\eqcomment{using \fref[plain]{eq:constraintalgebra:poisson-theta-complex-gen-ne}}\\
&= \sum_{c=1}^{n_c} \pangcomext{\rm I}{\rm II} \approx 0.
\eqnumber\label{eq:constraintalgebra:poisson-angle-complex-ne-gen}
\end{align*}
We see that the sum over all joined pairs of edges $(j_c,k_c)$ reduces
to only contributions from the pairs where $k_c = 1$ or $k_c = N$.

In the same way, we can generalize
$\poisson{C_\theta^{\rm I}}{C_z^{\rm I}}$ using
\fref[plain]{eq:constraintalgebra:poisson-theta-complex-gen-eq}, which
gives
\begin{align*}
&\nop \poisson{C_\theta^{\rm I}}{C_z^{\rm I}}\\
&=               \poisson{C^{\rm I}_\theta}{C^{\rm I}_z}_0 +
\sum_{c=1}^{n_c} \poisson{C^{\rm I}_\theta}{C^{\rm I}_z}_c\\
&= \pangcomint{\rm I} \\
&\nop +\sum_{c=1}^{n_c} \pangcomext{\rm I}{\rm I} \approx 0.
\eqnumber\label{eq:constraintalgebra:poisson-angle-complex-eq-gen}
\end{align*}

To generalize the Possion brackets of two complex (conjugate)
constraints to the case of multiple common edges, we first rewrite one
complex constraint to a sum of $\theta_j$ by the chain rule for
differentiation and obtain
\begin{align*}
&\nop \poisson{C^{\rm I}_z}{C^{\rm II}_z}\\
&= \poisson{\sum_{j=1}^N l_j\,e^{i\theta_j}}{\sum_{k=1}^M m_k\,e^{i\Phi_k}}\\
&= i\,\sum_{j=1}^N l_j\,e^{i\theta_j} \poisson{\theta_j}{C^{\rm II}_z}
  -i\,\sum_{k=1}^M m_k\,e^{i\Phi_k}   \poisson{\Phi_k}{C^{\rm I}_z}\\
&= i\,\sum_{j=1}^N l_j\,e^{i\theta_j} \sum_{c=1}^{n_c} \poisson{\theta_j}{C^{\rm II}_z}_c
  -i\,\sum_{k=1}^M m_k\,e^{i\Phi_k}   \sum_{c=1}^{n_c} \poisson{\Phi_k}{C^{\rm I}_z}\\
&= \sum_{c=1}^{n_c} \poisson{C^{\rm I}_z}{C^{\rm II}_z}_c
\eqnumber\label{eq:constraintalgebra:poisson-complex-complex-gen}
\end{align*}
The Poisson brackets where one of the constraints is a complex
conjugate and when both constraints are from the same polygon, are all
derived in an analogous way. For the explicit expression one must
insert the already calculated brackets with assumptions. See
\fref[plain]{eq:constraintalgebra:poisson-complex-complex-full} and
\fref{eq:constraintalgebra:poisson-complex-complexbar-full} for the
complete expressions.

\newpage
\subsection{Comparing for consistency}

Now we can do a consistency check with the results we have found. The
Hamiltonian can be written as the sum of all angular constraints. The
Poisson brackets of this sum of all angular constraints with one
complex constraint should thus match
expression~\fref{eq:constraintalgebra:poisson-hamil-complex}. Indeed
we find that they do match:
\begin{align*}
&\nop \poisson{H}{C_z^{\rm B}}\\
&= \sum_A \poisson{C_\theta^A}{C_z^B}\\
&= \poisson{C_\theta^B}{C_z^B} + \sum_{A\not=B} \poisson{C_\theta^A}{C_z^B}\\
&=-\left(  \frac{\sigma_N\,\gamma_1 + c_1\,\sigma_1\,\gamma_N}{s_1\,\sigma_1\,\sigma_N}
         -i\frac{\gamma_N}{\sigma_N}\right) \left(1 - e^{i\Phi_M}\right)\\
&\nop +\sum_{k=1}^M \left(1-e^{i\Phi_M}\right)\left(
     \frac{1            }{s_1\,\sigma_1}\,\delta_{k,1}
   + \frac{e^{-i\beta_1}}{s_1\,\sigma_M}\,\delta_{k,M}
  \right)\\
\intertext{%
  where we sum over all angular constraints, so all separate
  summations over common edges reduce to one summation over all edges
  of polygon B. This sum then reduces to the two terms $k = 1$ and
  $k = M$, which leaves us with%
}
&= \left(1 - e^{i\Phi_M}\right)\left[
   - \frac{\sigma_M\,\gamma_1 + c_1\,\sigma_1\,\gamma_M}{s_1\,\sigma_1\,\sigma_M}
   +i\frac{\gamma_M}{\sigma_M}
   + \frac{1            }{s_1\,\sigma_1}
   + \frac{e^{-i\beta_1}}{s_1\,\sigma_M}
  \right]\\
&= \left(1 - e^{i\Phi_M}\right)\left[
   -\frac{1}{s_1}\left(      \frac{\gamma_1 - 1}{\sigma_1}
                       +c_1\,\frac{\gamma_M - 1}{\sigma_M}\right)
   +i\frac{\gamma_M - 1}{\sigma_M}
  \right]\\
&= \left(-\frac{v_1 + c_1\,v_N}{s_1} + i\,v_N\right)
   \left(1 - e^{i\theta_N}\right)
\eqcomment{using $\dfrac{\gamma-1}{\sigma} = v$}\\
&= -\poisson{C_z^B}{H}
\eqcomment{inserting \fref[plain]{eq:constraintalgebra:poisson-hamil-complex}.}
\end{align*}
Note that we replaced $\theta$'s by $\Phi$'s when
inserting~\fref[plain]{eq:constraintalgebra:poisson-angle-complex-eq-gen}
and vice versa when rewriting the result back
to~\fref[plain]{eq:constraintalgebra:poisson-hamil-complex}, but these
are the same as both constraints describe the same polygon.

\newpage
\section{Summary}
\label{sec:constraintalgebra:summary}

The results of this chapter are summarized below. The same conventions
have been made and the sums are again over all pairs of common edges
between polygons $A$ and $B$. In the case that $A$ and $B$ are equal,
each pair will thus show up twice: if the pair $(j_c,k_c) = (x,y)$
exists, then also $(j_c,k_c) = (y,x)$.

\newlength{\oldjot}
\setlength{\oldjot}{\jot}
\setlength{\jot}{13pt}
\begin{subequations}\label{eq:constraintalgebra:poisson-full}
\begin{align*}
\poisson{C^A_\theta}{C^B_z}  &= \pangcomint{A}\,\delta_{A,B}\\
                             &\nop +\sum_{c=1}^{n_c}      \left(\rule{0pt}{0.8cm}\right.
                                \pangcomext{A}{B}
                                \left.\rule{0pt}{0.8cm}\right)
\eqnumber\label{eq:constraintalgebra:poisson-angle-complex-full}\\[8pt]
\poisson{C^A_z}{C^B_z}       &=     \sum_{c=1}^{n_c}      \left(\rule{0pt}{0.8cm}\right.
                                \pcomcomext{A}{B}{1.15cm} \left.\rule{0pt}{0.8cm}\right)
\eqnumber\label{eq:constraintalgebra:poisson-complex-complex-full}\\[8pt]
\poisson{C^A_z}{\bar{C}^B_z} &= \left(\pcomcbrint{A}\right)\,\delta_{A,B}\\
                             &\nop +\sum_{c=1}^{n_c}      \left(\rule{0pt}{0.8cm}\right.
                                \pcomcbrext{A}{B}{1.55cm} \left.\rule{0pt}{0.8cm}\right)
\eqnumber\label{eq:constraintalgebra:poisson-complex-complexbar-full}\\
\end{align*}
\end{subequations}
\setlength{\jot}{\oldjot}

\section{Interpretation}
\label{sec:constraintalgebra:interpretation}

We would like to interpret the Poisson bracket relations we have
found. The angular and complex constraints are first class constraints
and should correspond to some residual constraints and gauge freedom
left after the (partial) fixing of the constraints
$\mathcal{H}, \mathcal{H}^i$
in~\fref[plain]{eq:representation:ham-constr}
and~\fref{eq:representation:mom-constr}.

Intuitively, the residual constraint left from the Hamiltonian density
$\mathcal{H}$ is the sum of deficit angles at the vertices.
Superficially, this seems to match the angular constraints, but in
these the sum of deficit angle at a vertex is distributed among the
angular constraints of polygons that join at that vertex.

The complex constraints should be related to the residual gauge
freedom of Lorentz transformations of each polygon frame. With the
help of the dual graph formulation of
appendix~\fref[plain*]{chap:dualgraph}, we can try to see whether this
intuitive idea matches the explicit complex constraint definition. In
the dual graph, a Lorentz transformation of a polygon frame $f \in F$
amounts to shifting the corresponding vertex $\tilde{v} \in \tilde{V}$
in the dual graph by the action of the Lorentz transformation $L$ when
we choose $\tilde{v}$ to be at the origin $(1,0,0)$.

The boosts $\eta_i$ of edges along the polygon are encoded by half the
arc lengths of the geodesics emanating from $\tilde{v}$. Thus if we
make an infinitesimal Lorentz transformation, the change in boosts
will be half the ``length'' of the Lorentz transformation projected on
the direction of the outgoing edge (as seen in the dual graph). This
means that a Lorentz transformation of boost $\d\nu$ in the $x$
direction in the polygon frame should affect the boosts $\eta_i$ as
follows:
\begin{equation}
  \label{eq:constraintalgebra:lorentz-dn}
  \d\eta_i = \cos(\phi)\,\d\nu,
\end{equation}
with $\phi$ the angle that edge $i$ makes with the $x$ axis in the
dual graph. This angle is $\phi = \theta_i + \mfrac{\pi}{2}$, because
the direction of the boost is perpendicular to the direction of the
edge in the normal graph.

We can compare this to the gauge transformation generated by the
complex constraint. The action of that gauge transformation is given
by the Hamiltonian flow; the Poisson brackets of a quantity with the
constraint gives the action that the corresponding gauge
transformation has on that quantity. For the boosts we find
\begin{equation}
  \label{eq:constraintalgebra:poisson-dn}
  \d\eta_i
  = \poisson{\eta_i}{C_z}
  = -\mfrac{1}{2}\pder{C_z}{l_i}
  = -\mfrac{1}{2}e^{i\theta_i}
  = -\mfrac{1}{2}\left(\cos(\theta_i) + i\,\sin(\theta_i)\right).
\end{equation}
Comparing this with \fref{eq:constraintalgebra:lorentz-dn}, we see
that the imaginary part of $C_z$ generates Lorentz transformations in
the $x$ direction and the real part in the negative $y$ direction, at
least for the boost variables.

Unfortunately, the action on the length variables is much more
complicated. To compare the action on the lengths of the complex
constraint with the action of a Lorentz transformation, we can
consider the action of a Lorentz transformation of the polygon frame
geometrically and check whether that matches $\poisson{l_i}{C_z}$. In
the geometrical picture of the tessellation graph, one expects that an
(infinitesimal) Lorentz transformation acts on the variables as a
``rotation in space-time'' of the polygon plane, see
\fref{fig:polygon-rotation}. This changes the intersection lines with
its neighboring polygons, from which the action of the Lorentz
transformation on the lengths can be deduced. The explicit formulation
of this action is also quite complicated and we have not investigated
whether this geometrical picture matches the Poisson brackets.

%%% Local Variables: 
%%% mode: latex
%%% TeX-master: "thesis"
%%% End: 

%% file: conclusions.tex
\chapter{Conclusions}
\label{chap:conclusions}

We have looked at the polygon model as a way to describe $2\tp1$\ndash
dimensional gravity. This model makes use of the local flatness of
$2\tp1$D space-time in a remarkably efficient way: the polygon model
can be formulated as a finite dimensional system with a completely
explicit description of its evolution in terms of growth of edge
lengths and polygons undergoing transitions at a discrete set of
times. This makes the model very useful for exploring features of
gravity already in the classical context, e.g. by simulation.

In well known other descriptions of $2\tp1$\ndash dimensional gravity,
the system is defined more implicitly. For example in the ADM
formulation with York time, one must solve the conformal factor of the
metric from a elliptic partial differential equation. This has no
known explicit solution for genus $g>1$ \cite{Moncrief:1989dx}.
Witten's approach seems to use the frozen time picture, where the
dynamics cannot be recovered from the
system\cite{Moncrief:1990mk,Witten:1988hc}. These problems are not
present in the polygon model formulation.

There are however also some drawbacks of the formulation of the
polygon model. The polygons that together form a spatial slice, are of
finite size and internally flat by construction. In this construction
we have solved the field equations of the gauge constraints. The
finite dimensional system which is left, is (by definition) non-local
and still has constraints. This yields difficulties.

The problems of this non-locality are most clear in the Poisson
bracket structure of the constraints. The constraints for each polygon
are already non-local expressions and the Poisson brackets of
constraints between different polygons are complicated expressions,
which do not vanish off shell. Even stronger, they do not form a
proper Poisson algebra, because the brackets are not linear in the
constraints.

We have shown though, that the complete Poisson bracket structure can
be calculated explicitly and does vanish on shell. This was expected,
as the constraints should be a residue of the field constraints in the
unreduced canonical formulation, which are first class.

The features of the polygon model that allowed an explicit formulation
in the classical theory become problematic when we turn to the
question of quantization. First of all, a full description of a state
depends partially on a discrete graph structure, which changes under
transitions. This cannot easily be encoded in a quantum formulation.
Furthermore, there are inequality constraints which are coupled in a
very non-local way. This is due to the fact that the Hamiltonian is a
highly non-trivial function of the basic variables, which is only
defined when those inequalities are satisfied. Because of these
inequalities, we cannot lift the simple Poisson bracket structure
between those coupled variables to a quantum theory, as that would
violate these triangle inequalities.

We have shown that several ways of attacking the problem of finding a
consistent quantization run into one or more of the difficulties
above. Thus the question of whether the polygon model obtains a
discrete time and/or space spectrum after quantization seems to be
premature. It not only depends on whether one first quantizes and then
implements the constraints or vice versa, but more fundamentally on
whether a satisfactory quantum theory can be formulated in either
approach. If we first quantize, we face a complicated Poisson
structure which has to be implemented in the quantum theory. If we
first solve the constraints, we face the problem of finding an
explicit global solution to those.

Both approaches, although meeting some major obstacles, do not seem
completely hopeless. The dual graph formulation (as in
appendix~\fref[plain*]{chap:dualgraph}) gives a nice abstract
interpretation of the polygon model, which might be useful to reduce
the constraints in the classical theory in terms of holonomy-loop
variables. This dual graph formulation depends on the absence of
particles however. On the quantum theory side, one might consider
other variables than the canonical ones. This leaves the question of
whether this polygon model can be quantized to a difficult but open
one.

%%% Local Variables: 
%%% mode: latex
%%% TeX-master: "thesis"
%%% End: 

%% file: dualgraph.tex
\chapter{The dual graph}
\label{chap:dualgraph}

In this appendix we will present the dual graph of a tessellation in
the polygon model. For this we will first introduce the concept of a
dual graph in its general setting and also introduce hyperbolic space.
It turns out that we can assign a meaning to the dual graph of a
tessellation, such that it becomes a graph embedded in a hyperbolic
space: a surface with constant negative curvature.

\section{The dual of a planar graph}
\label{sec:dualgraph:dualgraph}

The concept of a dual graph is a very general one within graph theory
and has applications in a wide variety of fields. It does not add new
information to the graph, but allows one to see some features more
quickly by using a different representation of the same graph data.

To introduce the dual graph, we start with a graph $G = (V,E)$, where
$V$ is the set of its vertices and $E$ the set of edges. Each edge is
normally represented as a pair of vertices: the ones that it connects.
If we now embed our graph in the plane, the dual of that graph is
defined by associating a vertex to each region of the plane and
connect two vertices with an edge when their regions in the original
graph had an edge as a common boundary. One must however notice two
things: firstly, not every graph can be embedded in the plane without
intersecting edges, so this definition does not yield a dual graph for
each graph $(V,E)$. Secondly, a graph can have different possible
embeddings in the plane, resulting in different dual graphs; see
\fref[plains]{fig:dualgraph1}~and~\fref[plain*]{fig:dualgraph2} for an
example.
\begin{figure}
  \centering
  \begin{minipage}[t]{.45\textwidth}
    \centering
    \includegraphics{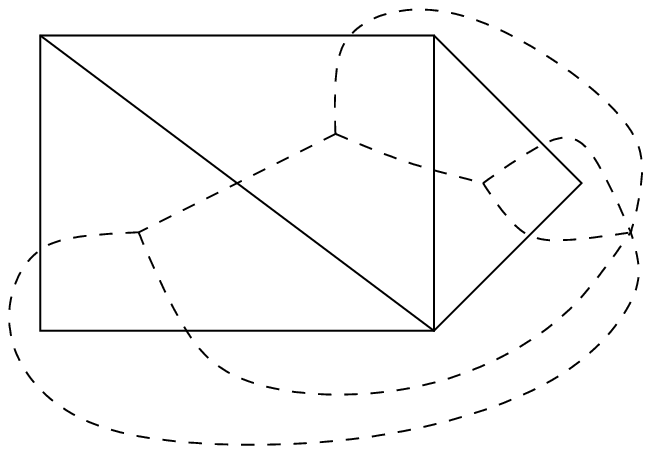}
    \caption{A graph embedded in the plane with its dual graph (dotted lines).}
    \label{fig:dualgraph1}
  \end{minipage}
  \hspace{1cm}
  \begin{minipage}[t]{.45\textwidth}
    \centering
    \includegraphics{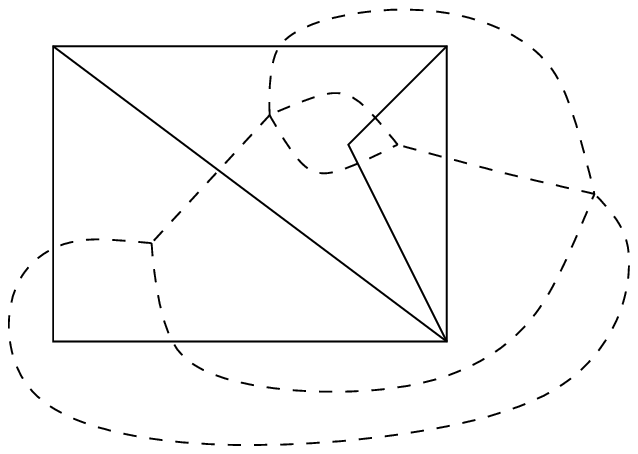}
    \caption{The same graph embedded differently with a different dual (dotted lines).}
    \label{fig:dualgraph2}
  \end{minipage}
\end{figure}

This non-uniqueness of the dual graph can be lifted if we add faces to
the graph information. We now write $G = (V,E,F)$ with $F$ a set of
faces and each edge also containing information about the pair of
faces it connects. We want this graph with added faces to uniquely
describe a graph on some genus $g$ surface. This means that we have an
orientation\footnote{%
  The orientation can only be defined locally in the case of a generic
  surface. However we only consider orientable surfaces, which allows
  us to extend this local orientation to a global one.} %
in the graph from the underlying manifold and that at each vertex and
face, we have an ordered list of the edges that ``connect'' to that
vertex or face.

With this added structure, we can assign a unique dual graph
$\tilde G$ to each graph $G = (V,E,F)$ by mapping vertices of $G$ to
faces of $\tilde G$ and faces of $G$ to vertices of $\tilde G$. The
edges are mapped to themselves with the pairs of vertices and faces
they connect interchanged. So we have
$\tilde G = (\tilde V,\tilde E,\tilde F) = (F,E,V)$.

Given this definition we can observe some facts. These only depend on
the graph structure, because at this point we have not yet added any
additional information to the graph, like edge lengths.

When we take the dual of the dual graph, we get back the original
graph:
\begin{equation}
  \widetilde{\tilde G} = G.
\end{equation}

From the geometrical picture of viewing the graph embedded in a
surface, we see that a graph $G$ and its dual $\tilde G$ represent
graphs on surfaces of the same homotopy class. When we look at the
Euler characteristic $\chi$ of the graph, we find the same result,
because we have
\begin{equation}
  \label{eq:dualgraph:eulerchar}
  \chi = 2 - 2\,g = \#V - \#E + \#F,
\end{equation}
where $g$ is the genus of the (orientable) surface.

\clearpage
\section{Hyperbolic space}
\label{sec:dualgraph:hyperbolic}

Hyperbolic space is a Riemannian manifold. Its most common definition
(and most practical here) is as follows. For $n \ge 2$, start with an
$(n\tp1)$\ndash dimensional Minkowski space, thus $\R^{n+1}$ with the
metric\footnote{%
  This can be interpreted as the metric on Minkowski space itself and
  the differential metric on the tangent space as Minkowski space is a
  vector space and thus isomorphic to its tangent space.} %
$\eta^{\mu\nu} = \text{diag}(-1,1,\ldots)$. Now $n$\ndash dimensional
hyperbolic space is defined as
\begin{equation}
  \label{eq:dualgraph:hyperbolicspace}
  \mathbb{H}^n = \{ x \in \R^{n+1} \,|\, \eta(x,x) = -1, \, x_0 > 0 \}
\end{equation}
with the metric induced from the underlying Minkowski space.

The space thus constructed is a true Riemannian manifold: the original
metric was a pseudo-Riemannian metric, but its restriction to
$\mathbb{H}^n$ is positive definite. Furthermore $\mathbb{H}^n$ is a
space with constant negative (sectional) curvature $-1$: at any point
$p$ and any $2$\ndash dimensional plane in the tangent space of $p$,
the scalar curvature is $-1$.

We will be concerned specifically with the case $n\!=\!2$: the
hyperbolic plane. The hyperbolic plane has some other well-known
representations, besides the
definition~\fref[plain]{eq:dualgraph:hyperbolicspace} given above.
One is that of the complex halfplane
\begin{align}
  \label{eq:dualgraph:complexplane}
  \{ x + i\,y \in \C \,|\, y>0 \}, &\qquad
  ds^2 = \frac{dx^2 + dy^2}{y^2},
\intertext{and the other is the unit disc}
  \label{eq:dualgraph:unitdisc}
  \{ (x,y) \in \R^2 \,|\, \norm{(x,y)}<1 \}, &\qquad
  ds^2 = \frac{dx^2 + dy^2}{1 - \norm{(x,y)}^2}.
\end{align}
These two models are known as the Poincar\'e plane and disc
respectively.

Hyperbolic space is a maximally symmetric Riemannian manifold. That
means that its isometry group is as large as can possibly be for a
Riemannian manifold, namely, of dimension $\frac{1}{2}n(n+1)$. We will
further investigate this symmetry from the viewpoint of our original
definition~\fref{eq:dualgraph:hyperbolicspace}. The symmetry group is
the identity component $SO^+(n,1)$ of the full Lorentz group
$O(n,1)$. The group $SO^+(n,1)$ consists of the time- and
space-orientation preserving Lorentz transformations on the Minkowski
space $\R^{n+1}$ and thus preserves the metric $\eta$ and the sign of
$x_0$.

This group $SO^+(n,1)$ can now be seen to have a group action on
$\mathbb{H}^n$: an element $L \in SO^+(n,1)$ defines a map
$L: \R^{n+1} \mapsto \R^{n+1}$ that can be restricted to a map
$L: \mathbb{H}^n \mapsto \mathbb{H}^n$, because $SO^+(n,1)$ precisely
preserves the quantities that define $\mathbb{H}^n$. This map $L$ then
also naturally defines the map
${\rm D}L: {\rm T} \mathbb{H}^n \mapsto {\rm T} \mathbb{H}^n$ on the
tangent bundle.

The elements $L \in SO^+(n,1)$ exactly correspond to all possible
isometries of $\mathbb{H}^n$. This is because the group $SO^+(n,1)$
has precisely $\frac{1}{2}n(n+1)$ degrees of freedom: $n$ degrees
correspond to translations of the hyperbolic plane and
\mbox{$\frac{1}{2}n(n-1)$} correspond to rotations around a fixed
origin.

Looking in detail at how $SO^+(n,1)$ acts on $\mathbb{H}^n$, we see
for $n = 2$ that pure Lorentz boosts $L_{x,y}(\eta)$ perform
translations in $\mathbb{H}^n$ by a distance of $\eta$ in the
$x,y$\ndash direction (locally around the ``origin'' $(1,0,0)$;
because of curvature the action is somewhat more complicated away from
the origin). The rotations $L_r(\alpha)$ perform a simple rotation
around the origin.

\section{Embedding the dual graph in $\mathbb{H}^2$}
\label{sec:dualgraph:embedding}

We have first constructed the dual graph of a tessellation. Let us now
assign geometrical data to this dual graph.

\begin{figure}
  \centering
  \begin{minipage}[t]{.45\textwidth}
    \center
    \includegraphics[scale=0.63]{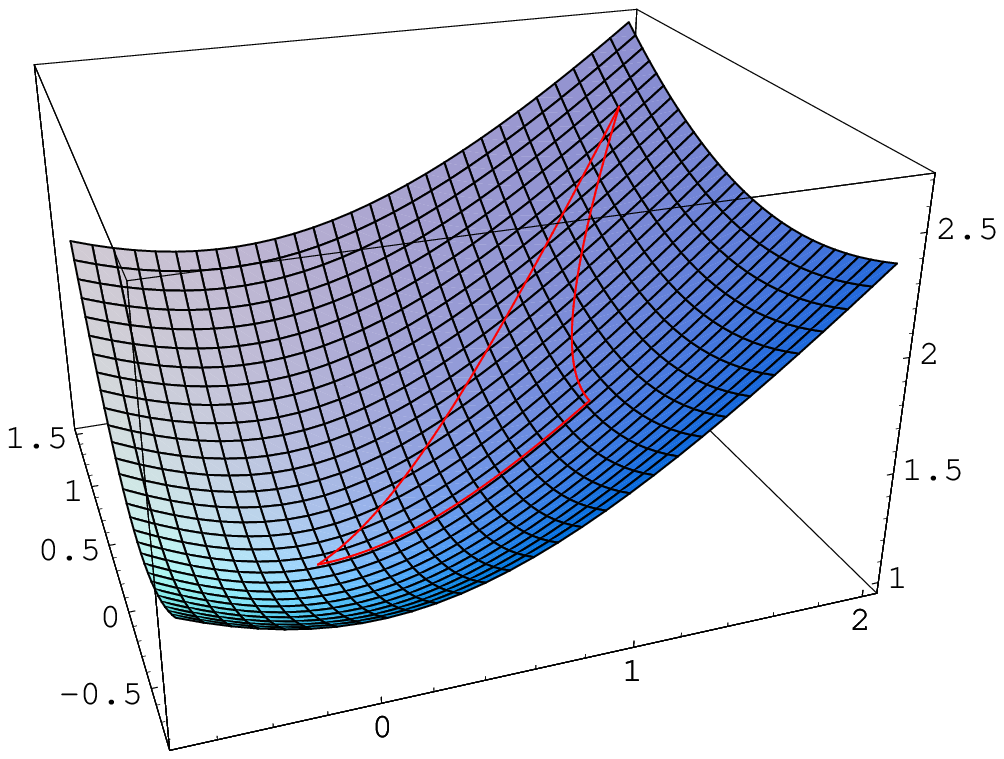}
    \caption{The hyperbolic surface $\mathbb{H}^2$ with red lines
      showing a triangle: the successive action of boosts and rotations,
      forming a closed holonomy.}
    \label{fig:hyperb-triangle}
  \end{minipage}
  \hspace{0.5cm}
  \begin{minipage}[t]{.40\textwidth}
    \center
    \includegraphics[scale=1.1]{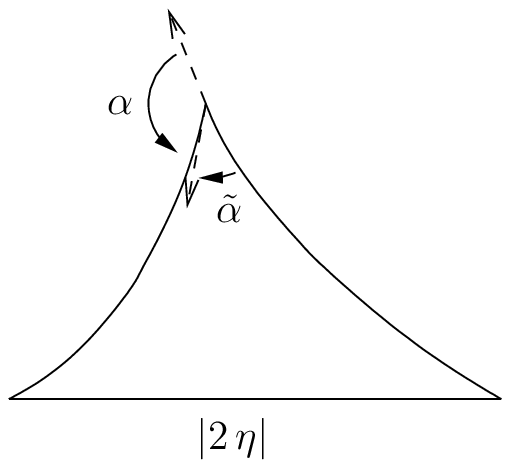}
    \caption{Angles of a hyperbolic triangle.}
    \label{fig:hyperb-triangle-params}
  \end{minipage}
\end{figure}

We will do so by looking at the path that a Lorentz transformation
maps out in hyperbolic space when it acts on $(1,0,0)$. Each Lorentz
transformation $L$ can be decomposed into pure rotations and boosts in
the following way:
\begin{equation}
  \label{eq:dualgraph:lorentz-decomp}
  L = L_r(\phi_1)\,L_x(\eta)\,L_r(\phi_2).
\end{equation}
This can be interpreted (in reverse order) as rotating the local frame
over an angle $\phi_1$ to the direction of the boost, then boost with
parameter $\eta$ and then rotate the boosted frame to the desired
orientation over an angle $\phi_2$. The continuous transformation of
$(1,0,0)$ under the one-parameter subgroup
\begin{equation}
  \label{eq:dualgraph:geodesic}
  L(t) = L_r(\phi_1)\,L_x(\eta\,t)\,L_r(\phi_2), \qquad t \in [0,1]
\end{equation}
will then be a geodesic in $\mathbb{H}^2$;
\fref{fig:hyperb-triangle} shows an example of such geodesics.

Now each edge in the dual graph corresponds to a Lorentz\footnote{%
  We can more generally assign to each edge a Poincar\'e
  transformation, by also taking into account the way coordinates are
  shifted after crossing an edge.} %
transformation $L$ of one frame to the other it connects. Actually, we
should consider oriented edges, because when we traverse the edge in
opposite direction, the Lorentz transformation is $L^{-1}$.  Thus each
edge can be viewed as an isometry of $\mathbb{H}^2$, which in turn can
be viewed as geodesics in $\mathbb{H}^2$.

A triangular polygon in the dual graph corresponds to a vertex in the
original graph. Thus traversing its three enclosing edges means going
around that vertex and the composition of corresponding Lorentz
transformations should be the identity. This means that, again viewing
these edges as geodesics in $\mathbb{H}^2$, they form a closed
triangle and we can thus naturally associate a graph structure with
it, see \fref{fig:hyperb-triangle}.

We can now try to interpret the geometrical data of this graph
embedded in $\mathbb{H}^2$: the dual graph lives in a Riemannian
manifold, so concepts like lengths and angles are well-defined. Some
subtleties will show up however, due to the fact that these geodesics
do not contain all information of the Lorentz transformations of the
edges. Each Lorentz transformation $L$ is fully characterized by the
way it acts as an isometry on $\mathbb{H}^2$, but if we only look at
its action on a single point, information is lost: the way it acts on
other points. This can be retrieved by not only giving the action of
$L$ on a single point, but also specifying its action on the tangent
space at that point.

When we now draw a triangle in the dual graph like in
\fref{fig:hyperb-triangle-params}, we find that each edge has geodesic
length $\abs{2\,\eta}$. The outer angles of the triangle are in direct
correspondence to the angles $\alpha_i$ at a vertex in the original
graph: they describe the rotation that we have to make, when we
``walk'' along a geodesic and turn to the direction of the next
geodesic. The inner angles $\tilde{\alpha}_i$ of the triangle are
related to these as
\begin{equation}
  \label{eq:dualgraph:dualangle}
  \tilde{\alpha}_i = \pi - \alpha_i.
\end{equation}
With this identification, we see that at each vertex in the dual
graph, we have a $2\,\pi$ neighborhood as expected:
\begin{equation}
  \label{eq:dualgraph:dual-ang-constraint}
  \sum_{i=1}^N \tilde{\alpha}_i =
  \sum_{i=1}^N \pi - \alpha_i = C_\theta = 2\,\pi
\end{equation}
by identification of the vertex in the dual graph with a polygon in
the original graph and using the angular constraint.

There is also an interpretation for the Hamiltonian
\fref[plain]{eq:polygon:hamiltonian} in the dual graph by considering the
area of all faces in the dual graph. We start from the fact that the
area of a triangle in hyperbolic space is determined by its inner
angles $\tilde{\alpha}_i$ only, as the deficit angle from $\pi$:
\begin{equation}
  \label{eq:dualgraph:area-triangle}
  A = \pi - \sum_{i=1}^3 \tilde{\alpha}_i.
\end{equation}
When we now sum the area of all triangles in the dual graph, we obtain
\begin{equation}
  \label{eq:dualgraph:dual-hamiltonian}
  A = \sum_{\tilde{f}\in\tilde{F}} \left(\pi - \sum_{i=1}^3 \tilde{\alpha}_i\right)
    = - \sum_{v \in V} \left(2\,\pi - \sum_{i=1}^3 \alpha_i\right)
    = - H.
\end{equation}

The boost parameters $2\,\eta$ cannot be interpreted as simply the
geodesic lengths: $\eta$ can be negative and in the geometric picture
of \fref{fig:hyperb-triangle-params} this corresponds to walking the
geodesic line in the opposite direction.

If we have a vertex where the signs of the three boosts are not the
same (called a `mixed vertex' in \cite{Kadar:2003ie}), this implies
that the interpretation of $\tilde\alpha_i = \pi - \alpha_i$ as the
inner angles of the triangle becomes problematic.

A way to solve this is to consider again the Lorentz transformations
corresponding to each edge. These can be decomposed as in
\fref{eq:dualgraph:lorentz-decomp}. In decomposing these we can always
choose $\eta$ positive: if it is negative, we can replace the
parameters by
$(\phi_1, \eta, \phi_2) \mapsto (\phi_1\pm\pi, -\eta, \phi_2\pm\pi)$.
In this way we get a graph with all edge lengths positive and the
oriented sum of angles at every point a multiple of $2\,\pi$. If we
translate this interpretation back to the original graph, it means
that we have reversed the orientation of the edges. This
interpretation thus moves the signs from the boosts onto the lengths,
which looks similar to the variables $x_i, p_j$ introduced in
\cite{'tHooft:1993nj}. This will affect the definition of the
Hamiltonian and the equations of motion, but we have not analyzed
further the implications.

See also the discussion of this problem of the interpretation of mixed
vertices in \cite[p.~47]{Kadar:2005phd}.

%%% Local Variables: 
%%% mode: latex
%%% TeX-master: "thesis"
%%% End: 

%% file: acknowledgements.tex
\chapter*{Acknowledgements}

First of all, I would like to thank my supervisor, Prof. Renate Loll
for giving me the opportunity to work on this interesting subject. She
guided me with my questions to the right places. I would especially
like to thank Zolt\'an K\'ad\'ar for the numerous discussions I had
with him and the detailed proofreading he did.

I also want to thank my fellow students in MG301 for the pleasant
atmosphere and the innumerable (work related) discussions. In
particular I want to express thanks to Gerben, Jan and Lotte for the
years of working together during our studies and Gerben for some
helpful discussions and ideas.

%%% Local Variables: 
%%% mode: latex
%%% TeX-master: "thesis"
%%% End: 